\documentclass[useAMS,usenatbib]{mn2e}
\usepackage[pdftex]{graphicx}

\def\aj{{AJ}}%
\def\araa{{ARA\&A}}%
\def\aap{{A\&A}}%
\def\aaps{{A\&AS}}%
\def\pasp{{PASP}}%
\def\mnras{{MNRAS}}%

\title[Galaxy Zoo Supernovae]{Galaxy Zoo Supernovae}
\author[Smith et al.]{A. M. Smith$^{1}\thanks{This publication has been made possible by the participation of more than 10,000 volunteers in the Galaxy Zoo Supernovae project (\texttt{http://supernova.galaxyzoo.org/authors}).}$$\thanks{E-mail: arfon.smith@astro.ox.ac.uk}$, S. Lynn$^{1}$, M. Sullivan$^{1}$\thanks{E-mail: sullivan@astro.ox.ac.uk}, C. J. Lintott$^{1}$, P. E. Nugent$^{2}$,\newauthor J. Botyanszki$^{2}$, M. Kasliwal$^{3}$, R. Quimby$^{3}$, S. P. Bamford $^{14}$, L. F. Fortson$^{15}$,\newauthor K. Schawinski$^{4,5,6}$,I. Hook$^{1,7}$, S. Blake$^{1}$, P. Podsiadlowski$^{1}$, J. J\"{o}nsson$^{1}$, A. Gal-Yam$^{8}$,\newauthor I. Arcavi$^{8}$, D. A. Howell$^{9,10}$, J. S. Bloom$^{11}$, J. Jacobsen$^{2}$, S. R. Kulkarni$^{3}$,\newauthor N. M. Law$^{12}$, E. O. Ofek$^{3,4}$, R. Walters$^{12}$\\\\
  $^{1}$Department of Physics (Astrophysics), University of Oxford, DWB, Keble Road, Oxford OX1 3RH, UK\\
  $^{2}$Computational Cosmology Center, Lawrence Berkeley National
  Laboratory, 1 Cyclotron Road, Berkeley, CA 94720, USA.\\
  $^{3}$Cahill Center for Astrophysics, California Institute of Technology, Pasadena, CA, 91125, USA\\
  $^{4}$Einstein Fellow\\
  $^{5}$Department of Physics, Yale University, New Haven, CT 06511, USA\\
  $^{6}$Yale Center for Astronomy and Astrophysics, Yale University, P.O. Box 208121, New Haven, CT 06520, USA\\
  $^{7}$INAF-Osservatorio di Roma, via Frascati 33, I-00040 Monteporzio Catone (Roma), Italy\\
  $^{8}$Department of Particle Physics and Astrophysics, Faculty of Physics, The Weizmann Institute of Science, Rehovot 76100, Israel\\
  $^{9}$Las Cumbres Observatory Global Telescope Network, 6740 Cortona Dr, Suite 102, Goleta, CA 93117\\
  $^{10}$University of California, Santa Barbara, Broida Hall, Mail Code 9530, Santa Barbara, CA 93106-9530, USA\\
  $^{11}$Department of Astronomy, University of California, Berkeley, CA 94720-3411, USA\\
  $^{12}$Dunlap Institute for Astronomy and Astrophysics, University of Toronto, 50 St. George Street, Toronto M5S 3H4, Ontario, Canada\\
  $^{13}$Caltech Optical Observatories, California Institute of Technology, Pasadena, CA 91125, USA\\
  $^{14}$School of Physics and Astronomy, University of Nottingham, University Park, Nottingham, NG7 2RD\\
  $^{15}$School of Physics and Astronomy, University of Minnesota,
  Minneapolis, MN 55455, USA}
\begin{document}

\maketitle

\begin{abstract}
  This paper presents the first results from a new citizen science
  project: Galaxy Zoo Supernovae.  This proof of concept project uses
  members of the public to identify supernova candidates from the
  latest generation of wide-field imaging transient surveys.  We
  describe the Galaxy Zoo Supernovae operations and scoring model, and
  demonstrate the effectiveness of this novel method using imaging
  data and transients from the Palomar Transient Factory (PTF).  We
  examine the results collected over the period April--July 2010,
  during which nearly 14,000 supernova candidates from PTF were
  classified by more than 2,500 individuals within a few hours of data
  collection.  We compare the transients selected by the citizen
  scientists to those identified by experienced PTF scanners, and find
  the agreement to be remarkable -- Galaxy Zoo Supernovae performs
  comparably to the PTF scanners, and identified as transients 93\% of
  the $\sim130$ spectroscopically confirmed SNe that PTF located
  during the trial period (with no false positive identifications).
  Further analysis shows that only a small fraction of the lowest
  signal-to-noise SN detections ($r>19.5$) are given low scores:
  Galaxy Zoo Supernovae correctly identifies all SNe with
  $\geq8\sigma$ detections in the PTF imaging data. The Galaxy Zoo
  Supernovae project has direct applicability to future transient
  searches such as the Large Synoptic Survey Telescope, by both
  rapidly identifying candidate transient events, and via the training
  and improvement of existing machine classifier algorithms.
\end{abstract}


\section{Introduction}
\label{sec:introduction}

Supernovae (SNe) have a profound influence upon many diverse areas of
astrophysics.  They are the key source of heavy elements in the
universe, driving cosmic chemical evolution.  Their energy input can
initiate episodes of star formation, and they are themselves the
product of the complex physics underlying the final stages of stellar
evolution.  The homogeneous nature of the thermonuclear Type Ia SNe
provides the most mature and direct probe of dark energy.  Despite
this importance in astrophysics, we understand surprisingly little
about the physics governing SN explosions.  Only the progenitors of
the core collapse Type IIP SNe have been directly identified: the
physical nature of other SN types remains uncertain \citep[for reviews
see][]{2000ARA&A..38..191H,2009ARA&A..47...63S}.  We remain ignorant
about many aspects of SN rates, light-curves, spectra, demographics,
and the dependence of these properties on environment, progenitor
composition, and explosion physics.
\\\\
In part, this is due to the historical difficulty and technical
challenges associated with locating SNe in the required numbers to
create statistically meaningful samples, particularly at low redshift
where high quality follow-up data can most easily be attained.  This
situation has changed with the availability of large format CCD
detectors.  Automated, wide-field transient searches on dedicated 1-2m
class telescopes and facilities are underway, typically observing
thousands of square degrees every few days
\citep[e.g.][]{2007PASA...24....1K,2009PASP..121.1395L}.  These
flux-limited `rolling searches' select transient events without regard
to host galaxy properties or type.
\\\\
This large amount of imaging data naturally generates its own
particular logistical challenges in dealing with the data flow, and
identifying transient astrophysical objects of interest in the data
(`candidates') for scientific study and analysis.  Of particular
importance is the rapid identification of new candidates once the
imaging data has been obtained and processed.  Though many aspects of
survey operations, such as image processing, can be efficiently
pipelined, the identification of new transient sources remains
challenging, with human operators (`scanners') invariably charged with
wading through new detections on a nightly basis. Though computer
algorithms can assist with identifying objects of interest in the
data, this scanning can still absorb a significant amount of
researcher time. A related issue is spectroscopic follow-up, a limited
resource that must be prioritised and allocated efficiently to the
detected candidates, with the absolute minimum of false candidates
observed.
\\\\
Two high-redshift SN searches highlight these challenges. The
Supernova Legacy Survey \citep[SNLS; e.g.][]{2006A&A...447...31A} used
the MegaCam instrument on the 3.6m Canada--France--Hawaii Telescope to
survey 4 deg$^{2}$ with a cadence of a few days.  Following automated
cuts on signal-to-noise and candidate shape, each square degree would
typically generate $\sim$200 candidates for each night of observation
\citep{2010AJ....140..518P}. Visual inspection would decrease this
number to $\sim$20 plausible real transients.  The Sloan Digital Sky
Survey-II Supernova Survey \citep[SDSS-SN;
e.g.][]{2008AJ....135..338F} used the SDSS 2.5m telescope to survey a
larger area of 300 deg$^{2}$, though to a shallower depth than SNLS
\citep{2008AJ....135..348S}. After the removal of moving (solar
system) objects, in the first season (3 month period), human scanners
viewed 3000--5000 objects each night spread over six scanners
($>$100,000 over the whole season). Although this number was radically
reduced in later seasons as more automated procedures were developed
($\sim$14,000 during season 2), the burden on human scanners was still
large \citep{2008AJ....135..348S}. With new wide-field transient
surveys generating many more candidates than these two surveys,
advances in both automated techniques and human scanning are clearly
required.
\\\\
This paper details a new method for sorting through SN candidates,
based upon the citizen science project `Galaxy Zoo'
\citep{2008MNRAS.389.1179L,2010arXiv1007.3265L}.  New candidate
transient events are uploaded to the Galaxy Zoo Supernovae website, and
are visually examined and classified by members of the public, guided
by a tutorial and associated decision tree.  Each candidate is
examined and classified by multiple people and given an average score,
with the candidates ranked and made available for further
investigation in real-time.  The advantages of this approach are
considerable.  First, the burden of candidate scanning is largely
removed from the science team running the survey.  Second, each
candidate is inspected multiple times (versus once by a scanner in
previous transient surveys), reducing the chances that the candidate
could be missed.  Third, with a large number of people scanning
candidates, more candidates can be examined in a shorter amount of
time -- and with the global Zooniverse (the parent project of Galaxy
Zoo) user base this can be done around the clock, regardless of the
local time zone the science team happens to be based in.  This speed
can even allow interesting candidates to be followed up on the same
night as that of the SNe discovery, of particular interest to quickly
evolving SNe or transient sources.  Fourth, the large number of human
classifications collected can be used to improve machine learning
algorithms for automated SNe classification.
\\\\
This paper reports the results from the early operations (over $\sim$3
month period) of this system.  In section \ref{sec:palom-trans-fact}, we
describe the Palomar Transient Factory, data from which were used in
the tests and running of Galaxy Zoo Supernovae.
Section \ref{sec:oper-galaxy-zoo} describes Galaxy Zoo Supernovae,
including the ranking system for candidates used by the citizen
science classifiers.  Section \ref{sec:results} has details of the tests
and first results of the Galaxy Zoo Supernovae operation.  We discuss
the future direction of this project in section \ref{sec:future}.

\section{The Palomar Transient Factory}
\label{sec:palom-trans-fact}

The Palomar Transient Factory (PTF) is a wide-field survey exploring
the optical transient sky.  The survey is built around the 48 inch
Samuel Oschin telescope at the Palomar Observatory, recently equipped
with the CFH12k mosaic camera (formerly at the Canada-France-Hawaii
Telescope) offering an 7.8 square degree field of view, and
robotised to allow remote and automated observations.  Observations
are mainly conducted using the Mould-$R$ filter.
\\\\
A full description of the operations of the PTF experiment can be
found in \citet{2009PASP..121.1395L}.  Of most relevance for SN
studies are the `5-day cadence' and `dynamical cadence' experiments,
each using $\sim40$ \% of the observing time.  The dynamic cadence
revisits survey fields on time-scales of 1 minute up to 5 days and is
particularly sensitive to rapid transient events (as well as longer
duration SNe), whereas the 5-day cadence is specifically targeted to
extra-galactic SN studies \citep{2009PASP..121.1334R}. Even in the 5
day cadence, images are typically taken in pairs separated in time by
about one hour.  This is to help identify moving objects (i.e.,
asteroids) in the imaging data, which might otherwise masquerade as
new transients.

\subsection{PTF real-time operations}
\label{sec:ptf-pipeline}

\begin{figure*}
\includegraphics [width=0.49\textwidth]{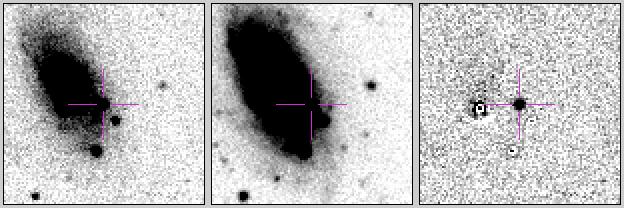}\hspace{0.01\textwidth}\includegraphics [width=0.49\textwidth]{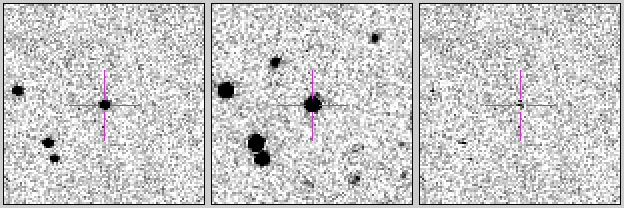}\vspace{0.01\textwidth}
\includegraphics [width=0.49\textwidth]{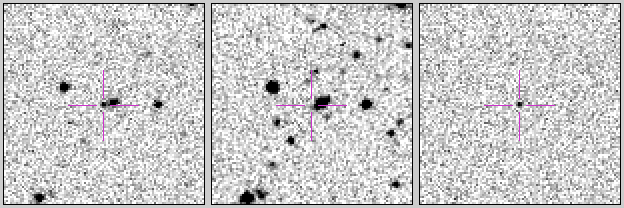}\hspace{0.01\textwidth}\includegraphics [width=0.49\textwidth]{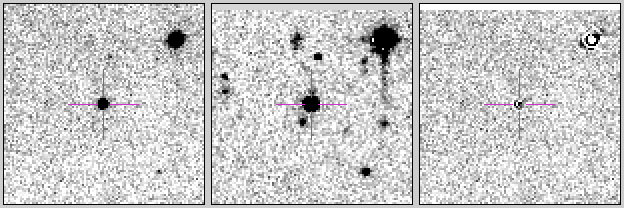}
\caption{Four example detection triplets from PTF, similar to those
  uploaded to Galaxy Zoo Supernovae. Each image is 100\arcsec\ on a
  side.  In each triplet, the panels show (from left to right) the
  most recent image containing the candidate SN light (the science
  image), the reference image generated from data from an earlier
  epoch with no SN candidate light, and the subtraction or difference
  image -- the science image minus the reference image -- with the SN
  candidate at the centre of the crosshairs. The two triplets on the
  left are real SNe, and were highly scored by the Zoo; the triplets
  on the right are not SNe and were given the lowest possible
  score.\label{fig:egtriplets} }
\end{figure*}

The PTF (near)-real-time search pipeline is hosted by the National
Energy Scientific Computing Center (NERSC) at the Lawrence Berkeley
National Laboratory (LBNL). After data is taken and transferred from
the Palomar observatory to NERSC, the pipeline generates new
subtraction images within an hour \citep{2010ATel.2600....1N},
subtracting an older, deep `reference' image from the new
observations.  The two images are photometrically matched using the
\textsc{hotpants}
program\footnote{\texttt{http://www.astro.washington.edu/users/becker/hotpants.html}},
an implementation of the \citet{2000A&AS..144..363A} algorithm.
Candidate transient events are then identified as $\geq5\sigma$
detections in the subtraction images using SExtractor
\citep{1996A&AS..117..393B}.  Fluxes and various other relevant
parameters are measured before storing all candidates in a database.
Each candidate is also `scored' (producing the PTF `real-bogus' value)
using a machine-learning algorithm (the `PTF robot') based on the
characteristics of the detection and previous history of the candidate
(Bloom et al., in prep.).  The vast majority ($\sim99.99$\%) of these
candidates are not real astrophysical transient events -- the search
algorithm is designed to be as inclusive as possible, with most of the
candidates rejected via simple cuts. These include: 
\begin{enumerate}
\item The ratio of both semi-major and semi-minor axes of the
  candidate shape to the seeing must be greater than 0.15 and less
  than 0.85, and the ratio of the FWHM of the candidate to the seeing
  must be greater than 0.5 and less than 2.0. These ensure that the
  candidate has a reasonable spatial extent when compared to the
  seeing,
\item In a 7 pixel by 7 pixel box placed on top of the candidate, the
  number of pixels deviating by more than 2$\sigma$ must be less than
  6, and the number deviating by more than 3$\sigma$ must be less than
  2.
\item Each candidate must be seen in at least one image taken in the
  previous 10 nights (including the night of detection), a constraint
  designed to remove fast moving solar system objects,
\item Candidates within 1\arcsec\ of previously located objects
  (excluding the previous 10 nights) are removed to avoid the repeated
  detection of (e.g.) AGN or variable stars.
\end{enumerate}
The effectiveness of these cuts means that a typical full night of PTF
observing will yield $\sim$100--500 (average $\sim$200) candidates
that survive these culls, which can then be further sorted using only
a short decision tree in Galaxy Zoo Supernovae.
\\\\
Though the ultimate aim is to make the human scanners redundant with a
fully automated machine-learning classification pipeline, at the
current time a substantial amount of human scanning is still required
to identify the good candidates (in part, this scanning can be used to
train machine-based methods).  Candidates are inspected visually by
human scanners in the PTF team, using a web interface to reject false
transient detections.  The human scanner can dynamically alter a set
of cuts to control the candidates that are shown for a given image,
including the signal-to-noise, shape parameters, the full-width
half-maximum (FWHM) of the candidate compared to the global image
value, and the output score from the machine classifier.  Based on the
cuts chosen, the scanner is presented with a series of detection
`triplets' -- each triplet contains three images showing the current
image of the field (containing SN light together with all other
objects), the historical or reference image of the same field (with no
SN light), and the difference between the two (which should contain
only the SN light). Examples of triplets are shown in
Fig.~\ref{fig:egtriplets}. The human scanner then decides, based on
his or her subjective (but informed) judgement, whether each of the
candidates presented is a real transient event, and if so marks that
candidate as either a SN-like transient or a variable star.
\\\\
The primary goal of Galaxy Zoo Supernovae in PTF is to initially
supplement, but perhaps ultimately replace, the role of the PTF human
scanners.  By presenting a transient candidate to a number of
different classifiers not only is the time of the PTF team freed to
spend on tasks not suitable for the general public, but the potential
of mis-classification of candidates due to individual human error is
significantly reduced.  The 5-day and dynamical cadence programs in
PTF collect data on every night of the year March to November (weather
permitting) and on each night 2--4 of the PTF team share the scanning
tasks, examining $\sim$500 candidates.  This not only requires several
person-hours of work, but the large number of classifications by a
small number of PTF-scanners is likely to contain errors, and this is
where the repeat-classification by Galaxy Zoo Supernovae volunteers
can help.
\\\\
The Galaxy Zoo Supernova project also has other aims. A longer-term
goal is to provide sufficient classification data for the training and
improvement of the PTF machine-learning classification algorithm. A
final consideration is to build expertise in the citizen science
community for future transient surveys, which of course generate many
more candidates than PTF, perhaps approaching thousands of genuine
candidates on a nightly basis.

\section{Galaxy Zoo Supernovae}
\label{sec:oper-galaxy-zoo}

\subsection{Description of a typical `Zoo'}
\label{sec:descr-typic-zoo}

The Galaxy Zoo Supernovae
website\footnote{\texttt{http://supernova.galaxyzoo.org/}} is built
using the Zooniverse\footnote{\texttt{http://zooniverse.org}}
Application Programming Interface (API) toolset.  The Zooniverse API
is the core software supporting the activities of all Zooniverse
citizen science projects.  Built originally for Galaxy Zoo 2, the
software is currently being used by six different projects. The
Zooniverse API is designed primarily as a tool for serving up a large
collection of `assets' (for example, images or video) to an interface,
and collecting back user-generated interactions with these assets.
\\\\
So that the project website can retain a high performance during
spikes of activity, Galaxy Zoo Supernovae is hosted on Amazon Web
Services which provides a virtualised machine environment that can
auto-scale in size based upon server load.  The site uses the Elastic
Compute Cloud\footnote{\texttt{http://aws.amazon.com/ec2}} (EC2) for
web/database servers and the Simple Storage
Service\footnote{\texttt{http://aws.amazon.com/s3}} (S3) for image
storage.
\\\\
Image assets are presented to volunteers of the website through custom
user interfaces, designed to aid the volunteer in classifying the
object.  For many projects this interface takes the form of a decision
tree which walks the volunteer through a number of questions
concerning the current image.  The interaction of the volunteer with
the website produces a set of `annotations' which together constitute
a `classification' of the asset.  These are stored for later analysis
or in the case of Galaxy Zoo Supernovae are scored in real-time to
change the behaviour of the website.

\subsection{Galaxy Zoo Supernovae website operations}
\label{sec:site-operations}

Similar in nature to the original Galaxy Zoo 2 interface, Galaxy Zoo
Supernovae is a classic example of a `Zoo'.  When a new highly-scored
candidate is located in the PTF pipeline, an image triplet
(Fig.~\ref{fig:egtriplets}) of the candidate is automatically
uploaded, together with a small amount of metadata, to the Galaxy Zoo
Supernovae API.  Upon upload, the image is saved to
Amazon S3 (a file hosting service) and registered with the website.
Finding new SNe is time critical and our method of automatically
registering new assets with the API means that classifiers are
inspecting SN candidates discovered just hours earlier.  The interface for Galaxy Zoo Supernovae presents
these candidate detection triplets (just as with the PTF human
scanners, $\S$~\ref{sec:ptf-pipeline}) together with a decision tree
of questions and answers designed to help classify each candidate (see
Fig.~\ref{fig:decision_tree}).  Fig.~\ref{fig:zoo_schematic} displays the typical flow in the system.
Once a candidate has been classified (see below) it is instantly
available to the PTF team through a private web interface.

\begin{figure}
\includegraphics [width=0.5\textwidth]{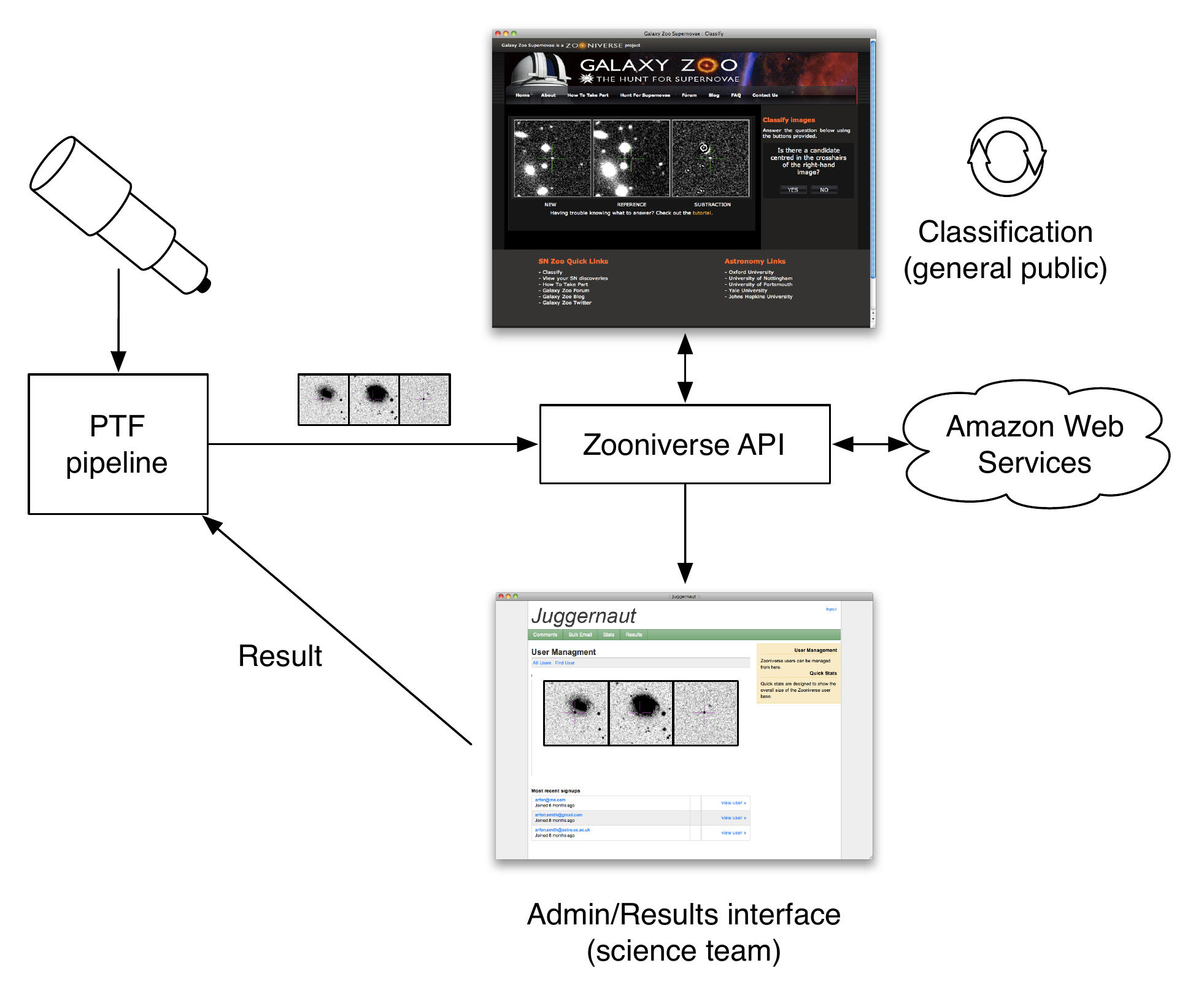}
\caption{A schematic showing the data acquisition and analysis in
  Galaxy Zoo Supernovae: Raw data is processed by the PTF pipeline,
  automatically uploaded to the API, presented, analysed and scored by
  the Zooniverse community and available for review by the PTF science
  team.\label{fig:zoo_schematic} }
\end{figure}

\subsection{Decision tree}
\label{sec:decision-tree}

\begin{figure}
\includegraphics [width=0.5\textwidth]{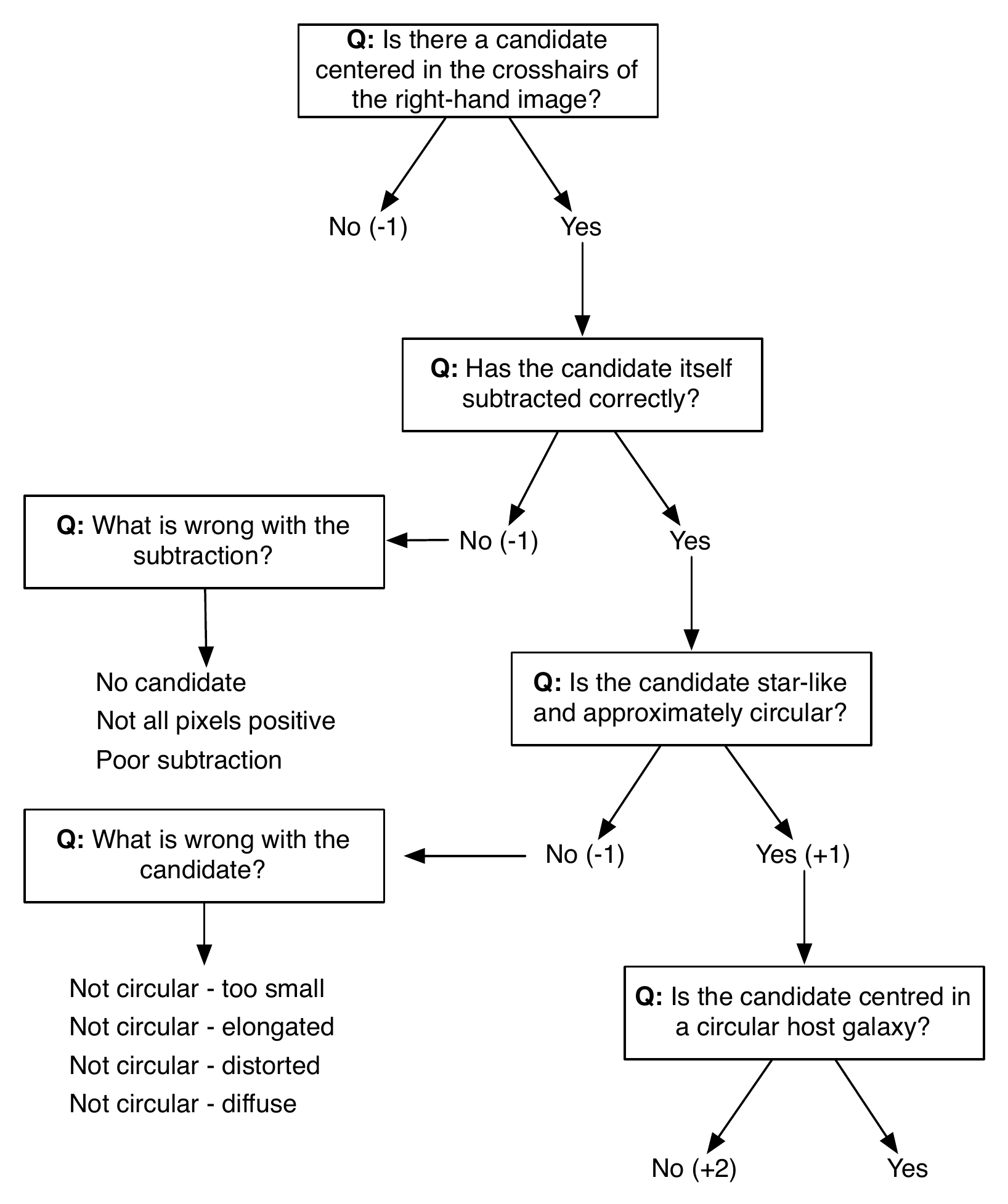}
\caption{The decision tree that a Galaxy Zoo Supernovae volunteer is
  presented with when classifying a candidate (see
  $\S$~\ref{sec:decision-tree}). The decision tree can end at a number
  of points. The scoring points in the decision tree are also shown.
  Both the path through the decision tree, and the cumulative score is
  recorded for later analysis.}
\label{fig:decision_tree}
\end{figure}

The decision tree developed to assist volunteers in classifying candidates
is described in Fig.~\ref{fig:decision_tree}.  This decision tree is
designed to remove as many false candidates as possible, without
losing real, scientifically interesting events.  In this respect the
decision tree is conservative in the candidates that are removed to
minimise the number of false negatives. The tree proceeds as follows:

\begin{enumerate}

\item \textit{Is there a candidate centered in the crosshairs of the
    right-hand image?}
  \\
  The PTF subtraction pipeline can occasionally undergo a failure and
  report (and therefore upload to the site) a `good' candidate that is
  actually an error in the processing.  This can be due to large
  (several pixel) mis-alignments of the two images being analysed,
  often localised in a particular part of the CCD where the
  astrometric solution fails.  Other sources of failure include
  saturated pixels or bleed trails from bright stars, or problems with
  the pipeline flat-fielding. The SExtractor detection algorithm can
  also sometimes detect a noise peak rather than a real transient.
  Though the basic cuts made by PTF remove most of these errors, on
  occasion they are ranked highly and uploaded to Galaxy Zoo
  Supernovae (emphasising the need for human classifiers).  Therefore,
  the first question in the decision tree is designed to remove such
  objects.  The right-hand image in the triplets in
  Fig.~\ref{fig:egtriplets} are the focus of this question.
  \\
\item \textit{Has the candidate itself subtracted correctly?}
\\
  Small mis-alignments between the reference and science image can
  result in image subtraction problems, usually indicated by a dipole
  of positive and negative pixels in the subtraction image. The cores
  of bright (but not saturated) stars can also mis-subtract, and
  result in `bullseye' patterns in the subtraction images. This
  question is designed to flag such candidates.
\\
\item \textit{Is the candidate star-like and approximately circular?}
\\
  This question is designed to remove unidentified cosmic rays, or
  diffuse/non-circular candidates which result from image subtraction
  problems. The volunteer is asked if the candidate looks like a round,
  symmetrical dot (star). Candidates that are very small (1--2 pixels,
  i.e., not PSF-like), elongated or otherwise distorted, or diffuse
  would trigger a negative response to this question.
\\
\item \textit{Is the candidate centered in a circular host galaxy?}
  \\
  The final question is more subjective, and is designed to categorise
  real astrophysical transients into two broad categories. Many of the
  transients which PTF detects are variable stars lying within our own
  galaxy, which are of interest to a different set of science users
  than extra-galactic transients. Variable star transients will appear
  to lie in `hosts' that are circular (as they are stars), and will
  also appear to be located in the centre of these hosts. By contrast, SNe
  will either have no host galaxy, or will lie (probably off-centre)
  in a large diffuse host galaxy.  This question therefore broadly
  splits the real transients into variable stellar transients, and
  SNe.  Most SNe that do happen to lie in the centres of their host
  galaxies will not be categorised as variable stars -- the question
  also requires the `host galaxy' to be circular.

\end{enumerate}
A full tutorial is available to new volunteers of the website to illustrate
the different questions using real PTF data.

\subsection{Asset scoring and priority}
\label{sec:scoring}

Once a volunteer has examined a candidate, their response is converted into
a score, $S$, as follows.
\begin{itemize}
\item The initial score is zero.
\item If a classifier answers negatively any question up to and
  including `Is the candidate star-like and approximately circular',
  the candidate is given a score of -1.
\item If a classifier instead answers positively up to that question,
  then the candidate is given a score of +1.
\item If the classifier then also marks the candidate as not centred
  in a circular host, then the candidate gains an \textit{additional}
  score of 2.
\end{itemize}
The structure of the decision and scoring of the questions means that
candidates can only end up with a score of -1, 1 or 3 from each
classification, with the most promising SN candidates scored 3. As
each new classification is received, the arithmetic mean score
($S_{\mathrm{ave}}$) of the candidate is recalculated. Candidates
which are not astrophysically interesting tend to have
$S_{\mathrm{ave}}<0$ (i.e., most volunteers scored them a `$-1$').
Astrophysical transients typically have $S_{\mathrm{ave}}>0$, and SNe
tend to have $S_{\mathrm{ave}}>1$ (i.e., most volunteers scored them a
`3').
\\\\
The asset prioritisation system is adjusted after each classification
is received, and operates to prioritise the best SN candidates (i.e.,
the order in which the candidates are shown to classifiers). When new
candidates are uploaded to the website, they are initially prioritised
based upon i) a score supplied by the PTF pipeline, and ii) the age of
the candidate (the newest uploads are shown first).  The PTF `real
bogus' value ($\S$~\ref{sec:ptf-pipeline}) is calculated by the PTF
pipeline for all candidates and gives an indication of the likelihood
that a candidate is a real transient. This value is only used to
determine the order in which candidates are shown and is not used in
the final ranking.
\\\\

Studies of results from early (`beta') versions of Galaxy Zoo Supernovae 
have allowed us to optimise the asset prioritisation to reduce the time taken to
identify candidates. We divide candidates in to four categories:
\begin{enumerate}
\item \emph{Unseen} -- Candidates which have 3 or fewer
  classifications.
\item \emph{Bulk} -- Candidates which have been classified between 3
  and 10 times.
\item \emph{Stragglers} -- Candidates which have been classified more
  than 10 times, but which do not have a `definitive'
  $S_{\mathrm{ave}}$ (i.e., those with  $0.0 < S_{\mathrm{ave}} < 1.7$).
\item \emph{Done} -- Candidates which have been classified more than
  10 times and which have $S_{\mathrm{ave}}<0.0$ or
  $S_{\mathrm{ave}}>1.7$, and candidates which have been classified
  more than 20 times.
\end{enumerate}
Candidates in the `unseen' category are given absolute precedence over
all others in an aim to get an initial understanding of the quality of
the candidate; they are shown in order of upload time followed by the
real-bogus score. Once these are completed, the `bulk' and `straggler'
candidate classes have equal priority. We select randomly between the
two classes, choosing the newest candidate with the highest score from
each group -- as a candidate begins to receive `positive'
classifications (i.e., $S$ of 1 or 2) then it is prioritised above any
others thus allowing rapid identification of the most interesting
targets.
\\\\
The choice of 10 classifications as the first point at which a
candidate can be considered classified is a compromise between the
robustness of the classification and speed. Clearly, the greater the
number of classifications required for each candidate the slower the
classification process proceeds; yet the process must be robust
against both user mistakes (i.e., clicking the wrong button) and
misunderstanding.
\\\\
The aim is to both quickly classify the best, high-scoring candidates
(which will rapidly exceed the $S_{\mathrm{ave}}=1.7$ threshold after
10 classifications), and to remove the worst candidates (which will
remain below $S_{\mathrm{ave}}=0.0$).  More ambiguous candidates can
then obtain up to 10 extra classifications before completion. The
process continues until a target has received enough classification
scores that it is considered `done', at which point it is removed from
the pool of available candidates.  Our simulations based on the beta
versions indicated that this scheme is 2--3 times faster at
classifying than just using a random order.

\subsection{Communication of results}
\label{sec:comm-results}

The science of Galaxy Zoo Supernovae relies on new candidates being
classified rapidly, and those classifications then being easily
accessible to the science team.

\subsubsection{Science dashboard}
\label{sec:science_dashboard}

A key part of the Galaxy Zoo Supernovae website is a science
`dashboard' for the PTF team.  The science dashboard provides basic
statistics on the number of candidate uploads, classifications and
volunteers versus time as well as a more in-depth breakdown of the
classification history for a candidate or individual.
\\\\
Custom views have been created which break down a score ranked list of
candidates for each day and week allowing observing teams to use these
rankings to help in the identification of good candidates for follow
up observations.  Candidates already identified as PTF transients show
the PTF identifier on the science dashboard and a link is also provided to
allow the science team to easily mark a highly-ranked candidate from
the Zoo in the PTF database.

\subsubsection{Candidate alerts}
\label{sec:cand_alerts}

In order to improve the rate at which objects are classified, an
automated alert system that monitors the number of candidates being
uploaded to the website is used.  Should the number of unclassified
candidates reach a threshold, the website sends an automated `alert'
to Galaxy Zoo Supernovae subscribers.  These (email) alerts are
usually sent out once per day, coinciding with the end of a night's
candidates being uploaded from NERSC, and usually result in the full
complement of candidates being classified within a few hours.

\subsubsection{`My Supernovae'}
\label{sec:my_sn}
Providing feedback to the Galaxy Zoo Supernovae community is a vital
part of the overall website experience to encourage volunteers to
return to the website.  This is partly done using forums and blogs
where scientists can comment on individual events classified by the
zoo. In addition, each volunteer can view a history of the candidates
that they have classified on their `My Supernovae' page.
\\\\
The `My Supernovae' (MySN) page displays the candidate triplets.
Those which have been observed are overlaid with a small symbol identifying the candidate as a SN, variable star, or asteroid.  Clicking on one of the candidates also
allows the volunteer to see the average rating across all
classifications, the number of classifiers and whether the candidate
was selected for followup by the PTF team.  PTF observers are
encouraged to leave comments on the science dashboard that the
classifiers can also see on their MySN page.

\section{Results}
\label{sec:results}

\begin{figure*}
\includegraphics[width=0.99\textwidth]{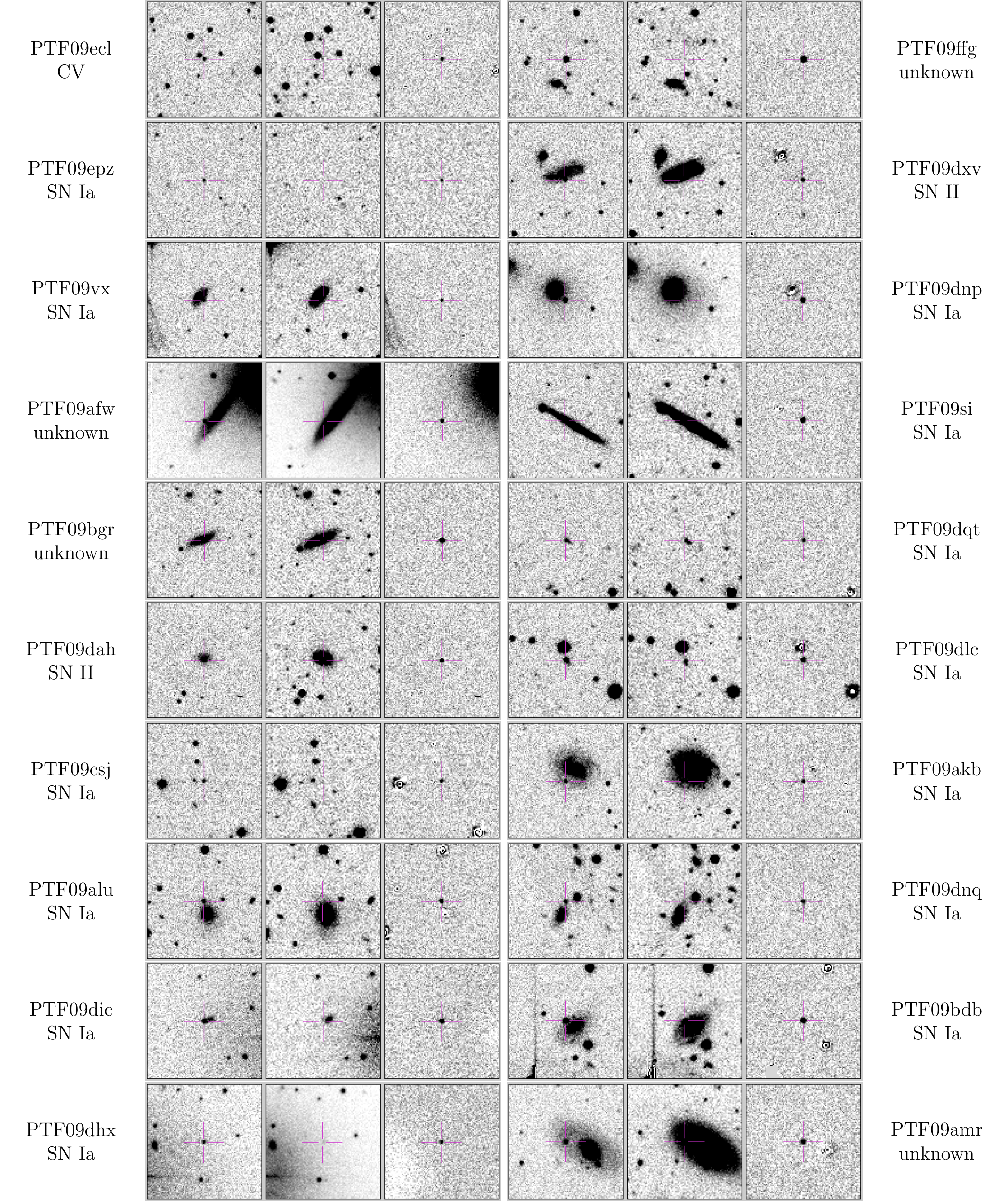}
\caption{A montage of the 20 highest ranked PTF candidates from the
  October testing of the website. Each set of three images shows, from
  left to right, the new image, the reference image, and the
  subtraction image. The position of the candidate is shown in each
  panel by the crosshairs. The candidate name and the spectroscopic
  type from the WHT (where available) are also
  shown.\label{fig:montage}}
\end{figure*}

\begin{figure}
\includegraphics [width=0.5\textwidth]{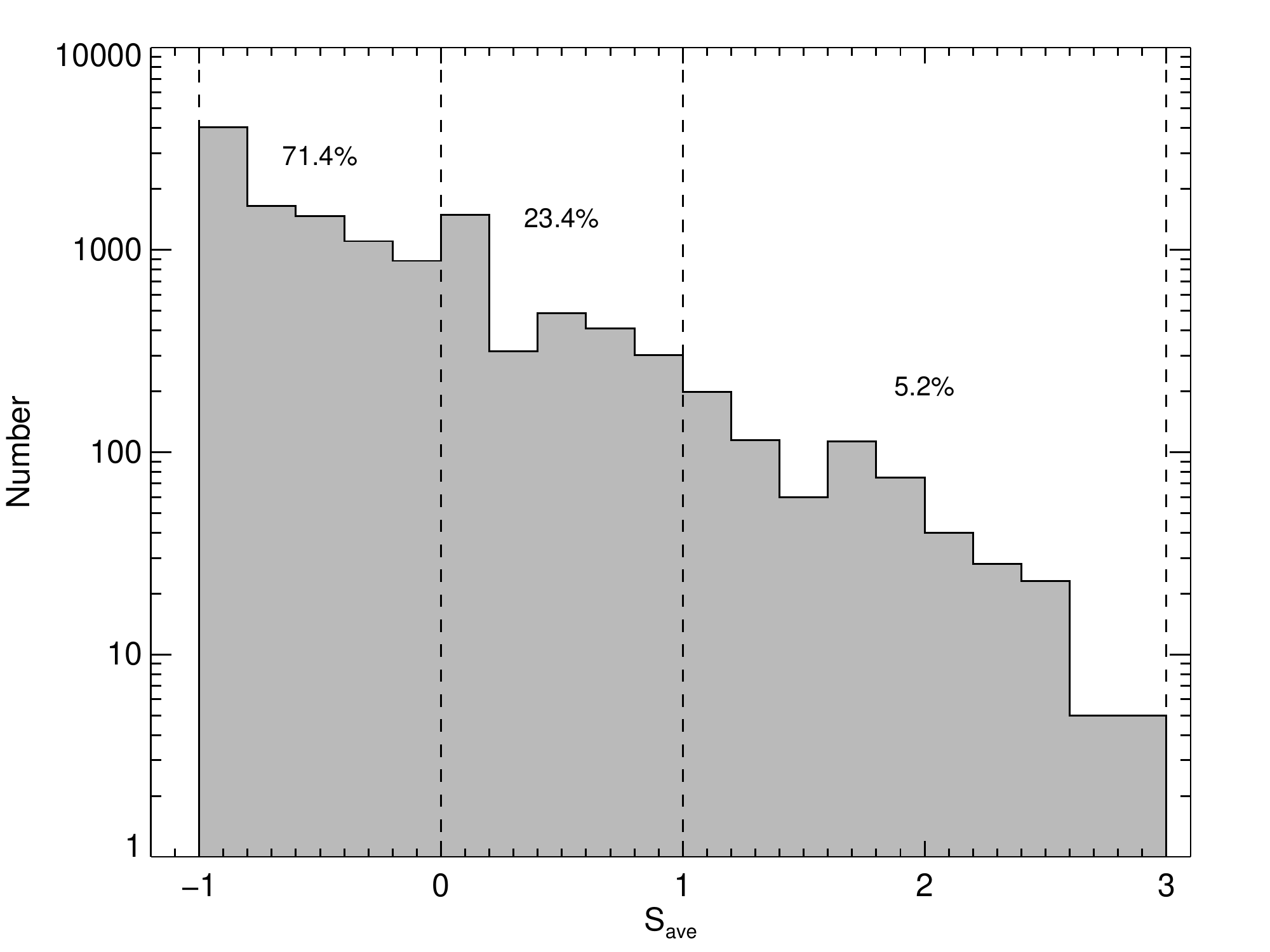}
\caption{The distribution of all of the scores ($S_{\mathrm{ave}}$)
  for all of the PTF candidates classified by Galaxy Zoo Supernovae
  between April 2010--July 2010. $\simeq13,900$ candidates were
  classified. The bulk of these -- $\simeq70 \%$ -- were classified as
  not astrophysically real by the zoo ($S_{\mathrm{ave}}<0$). Only 1
  in 20 candidates were identified as likely SN events.
  \label{fig:scoredist}}
\end{figure}

Galaxy Zoo Supernovae was first trialled on two specific occasions
supporting PTF spectroscopic follow-up observations at the 4.2m
William Herschel Telescope (WHT), in August 2009 and October 2009.
The selection of the candidates observed by WHT was guided by the Zoo
results, with a particular emphasis on comparing the classifications
produced by Galaxy Zoo Supernova with those produced by PTF human
scanners working on the same data. The top 20 scored candidates from
this initial trial run of Galaxy Zoo Supernovae are shown in
Fig.~\ref{fig:montage}. Sixteen of these candidates were observed by
WHT; 15 were confirmed as SNe, with 1 cataclysmic variable.
\\\\
Since April 2010, Galaxy Zoo Supernovae has been running full-time on
PTF candidates, and by July 15th 2010 had classified $\simeq13,900$ SN
candidates at the rate of several hundred candidates per observing
night. In all but the earliest weeks of the project, all submitted
candidates were classified by the zoo. This classified sample forms
the basis of our analysis in this section.  A distribution of the
scores ($S_{\mathrm{ave}}$) for all of these candidates can be found
in Fig.~\ref{fig:scoredist}. The bulk of the candidates uploaded are
classified as likely not astrophysically real events, and correspond
to subtraction artefacts or other reduction problems. This is
indicative of the conservative cuts that are made in the PTF pipeline
to avoid losing real SN events for follow-up, and highlights the
currently essential requirement for visual inspection of the pipeline
candidates.

\subsection{Comparison with professional classifiers}
\label{sec:comp-with-prof}

The performance of the public at classifying candidates can be gauged
by comparing with the classifications the PTF team assigned to the
same objects. The PTF team broadly classify objects into 4 visual
categories: not interesting (not assigned a type), asteroids, variable
stars, and transients (such as SNe). Asteroids are not screened for by
Galaxy Zoo Supernovae -- only one image is uploaded for each PTF
candidate, which clearly cannot be used to distinguish moving objects.
Asteroids are typically removed from the candidate list prior to
upload by insisting on two separate detections of a candidate within
1\arcsec\ of each other, though this process is not perfect,
particularly with slow moving asteroids where the apparent motion can
be only a few arcseconds a day.
\\
\begin{figure}
\includegraphics [width=0.5\textwidth]{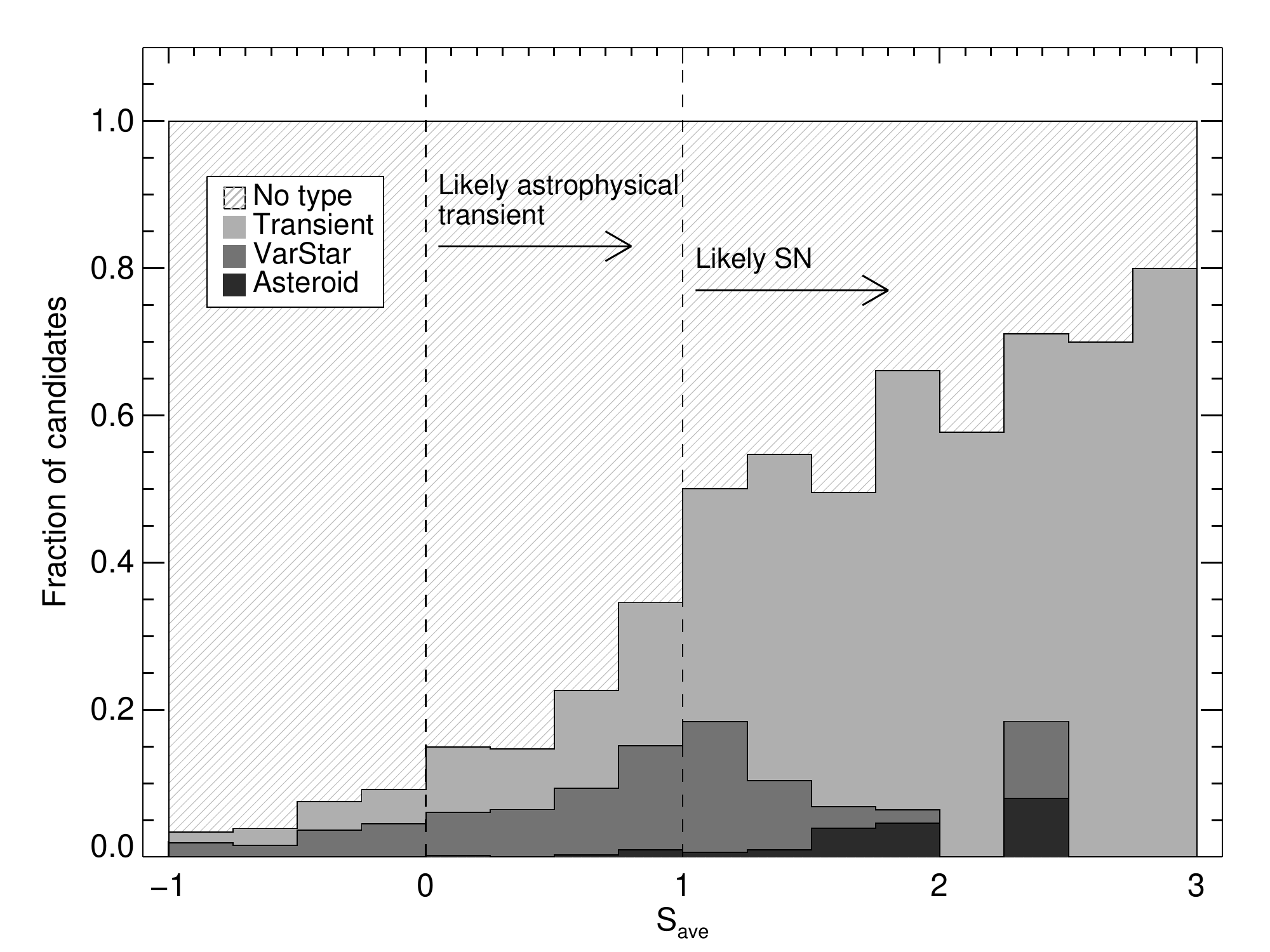}
\caption{A break down of the classifications collected during
  operations of Galaxy Zoo Supernovae. The bars show the distribution
  of candidate types (as determined by the PTF team) for a given zoo
  score ($S_{\mathrm{ave}}$).  The PTF team potentially assign a
  classification of asteroid, variable star, or transient to each zoo
  candidate -- objects without a PTF classification are deemed not to
  be interesting. Galaxy Zoo Supernovae is not designed to flag moving
  objects, which are largely removed before upload. Note that not all
  variable stars will be saved to the PTF database, so this category
  is likely highly incomplete (and the variables stars will appear as
  ``No type'').\label{fig:breakdown}}
\end{figure}

\noindent
To illustrate the performance of Galaxy Zoo Supernovae, we split the
candidates by their PTF assigned categories and calculate the fraction
in each category as a function of $S_{\mathrm{ave}}$.
Fig.~\ref{fig:breakdown} is a stacked box plot of the results. At low
scores practically all candidates are those which the PTF team decide
are not interesting: these will include poor subtractions,
artefacts/cosmic rays, etc.. As $S_{\mathrm{ave}}$ increases we see a
steady rise in the number of both variable star candidates and
transients. By a score of around $1.4$, variable stars are no longer
selected, and instead the majority of the candidates are SN-like
transients.
\\\\
A number of caveats should be borne in mind when examining this plot.
The first is that not all variable stars identified by Galaxy Zoo
Supernovae will be assigned that type by the PTF scanners. As the
primary goal of PTF is the study of explosive transients, variable
stars are frequently not recorded in the PTF catalogue (i.e., they
will be assigned ``No type'' in Fig.~\ref{fig:breakdown}). The second
caveat is that each PTF candidate is potentially observed many times
over a period of several weeks over many epochs, yet should only be
uploaded to Galaxy Zoo Supernovae once. If there is some problem with
the particular epoch that is uploaded to the zoo (a poor image
subtraction, or poor seeing conditions), then a real astrophysical
event may be poorly scored by the zoo on that epoch. However, that
candidate may potentially be saved by a human scanner based on an
image from a different epoch.  Thus real transient events can
occasionally be poorly scored by the zoo if the uploaded image is of
poor quality; this is the case for some of the real transients that
scored $S_{\mathrm{ave}}<0$. Finally, it is important to note that the
true nature of many of the candidates remains unknown, and the
comparison drawn here is between the zoo selection and that of a
subjective (though experienced) expert opinion.
\\\\
Figure~\ref{fig:breakdown} demonstrates that Galaxy Zoo Supernovae is
capable of prioritising good candidates, and that the highest ranked
candidates are likely to be SNe rather than variable stars.  The
candidates which were classified as asteroids in the Galaxy Zoo
Supernovae sample are given a relatively high score by the zoo
volunteers -- they typically mimic high-quality `hostless' transient
events.
\\\\


\begin{figure}
\includegraphics [width=0.5\textwidth]{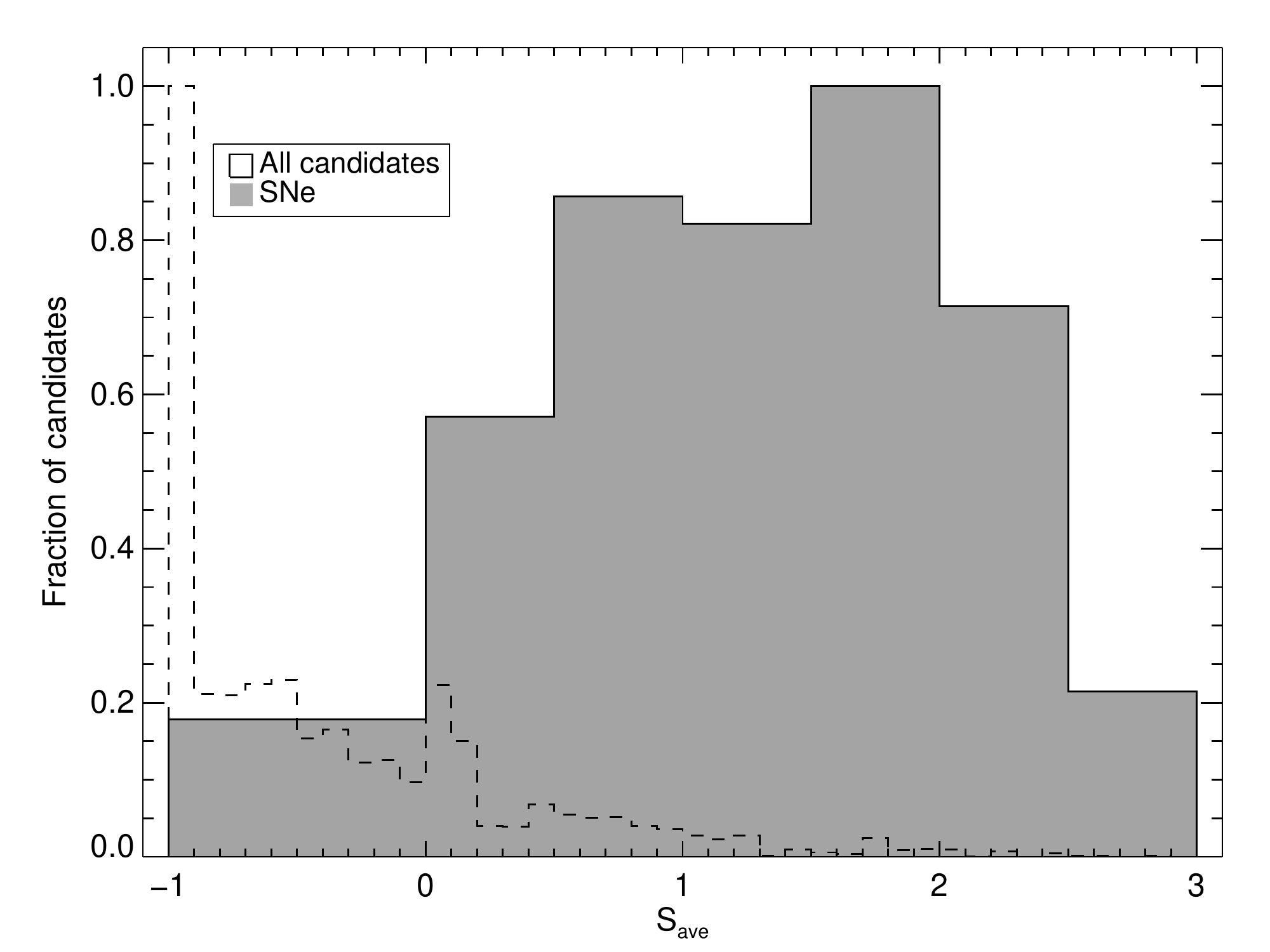}
\caption{A break down of the scores ($S_{\mathrm{ave}}$) for the 140
  known SNe identified via PTF follow-up spectroscopy (grey
  histogram). For reference, the distribution of the
  $S_{\mathrm{ave}}$ measures for all the objects is shown as the open
  histogram. These classifications were collected during April--July
  2010.\label{fig:snscores}}
\end{figure}

Some of the Galaxy Zoo Supernovae classified candidates were observed
spectroscopically by the PTF collaboration, as well as candidates
identified by other techniques. We examine the $S_{\mathrm{ave}}$
distribution for these $\sim140$ spectroscopically confirmed SNe
(Fig.~\ref{fig:snscores}), equivalent to $\sim$5--6 full nights of
4m-class telescope time (spread over 10 actual nights with a mix of
screening and follow-up of previously confirmed transients).
Approximately 93\% of these SNe gathered by PTF over April--July 2010
were highly-scored ($S_{\mathrm{ave}}>0$) by the zoo (60\% have
$S_{\mathrm{ave}}>1$), and real SNe with a $S_{\mathrm{ave}}<0$
comprise only 0.1\% of all zoo objects scored with
$S_{\mathrm{ave}}<0$. Though this may represent a slightly biased test
(low-scored candidates are less likely to be followed
spectroscopically), there are other techniques for screening
candidates within PTF that complement the zoo that partially mitigate
this bias. These include the PTF robot ($\S$~\ref{sec:ptf-pipeline})
and some human scanning effort. It is encouraging that, to the degree
that we can test it, the zoo is capable of selecting the best SNe from
the PTF dataset. We also note that no highly-scored zoo candidate that
was observed spectroscopically turned out not to be a SN.
\\\\
Note that the majority of the highly-scored Galaxy Zoo Supernovae
remain unobserved spectroscopically, particularly at the fainter end
of the candidate brightness. Some of these candidates are asteroids
(or more accurately, objects that only appear on a single night of
data).  Others are probably real SNe for which there was insufficient
follow-up time available.

\subsection{Effect of candidate brightness}
\label{sec:mageffect}

\begin{figure*}
\includegraphics[width=0.495\textwidth]{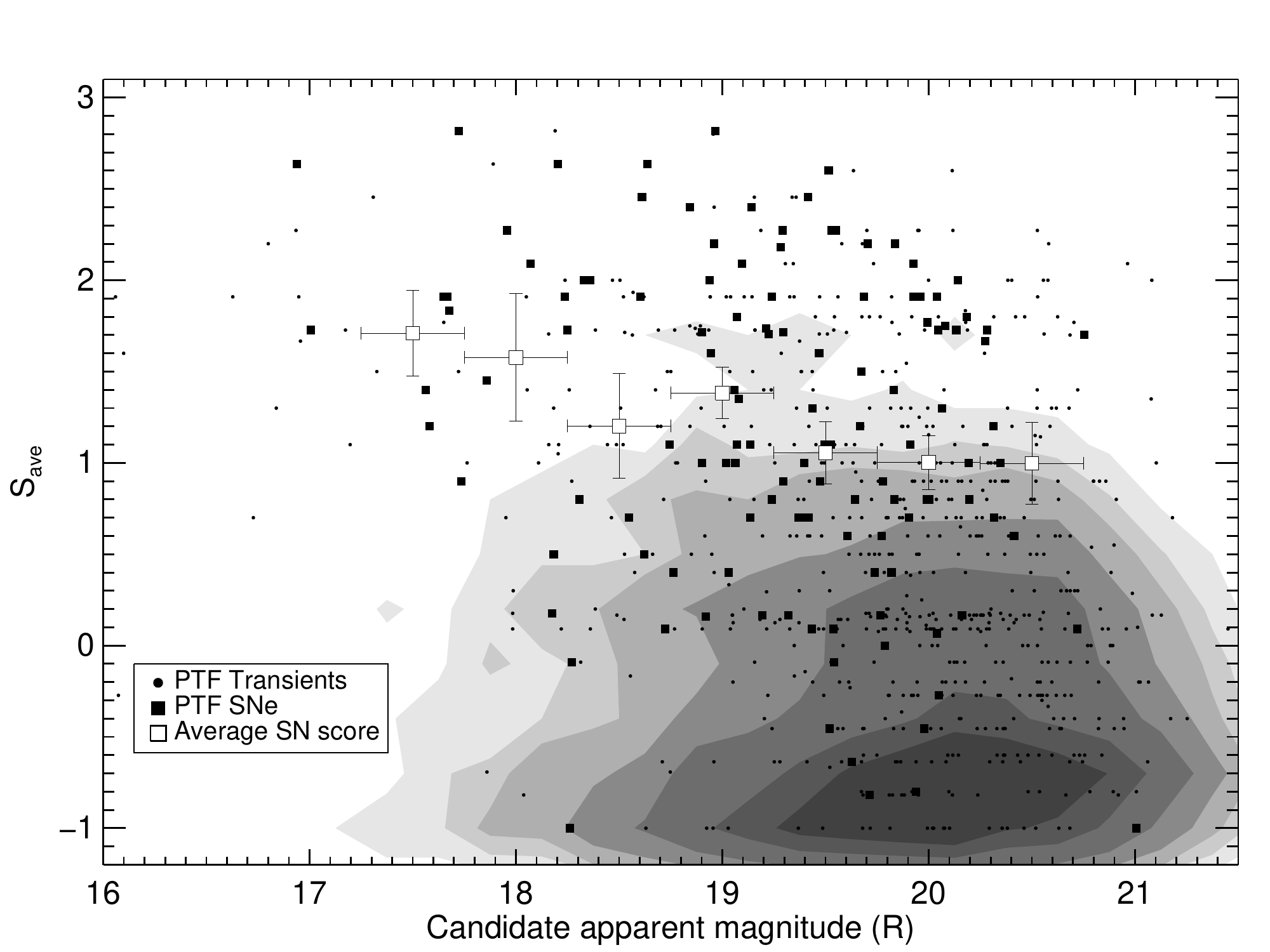}
\includegraphics[width=0.495\textwidth]{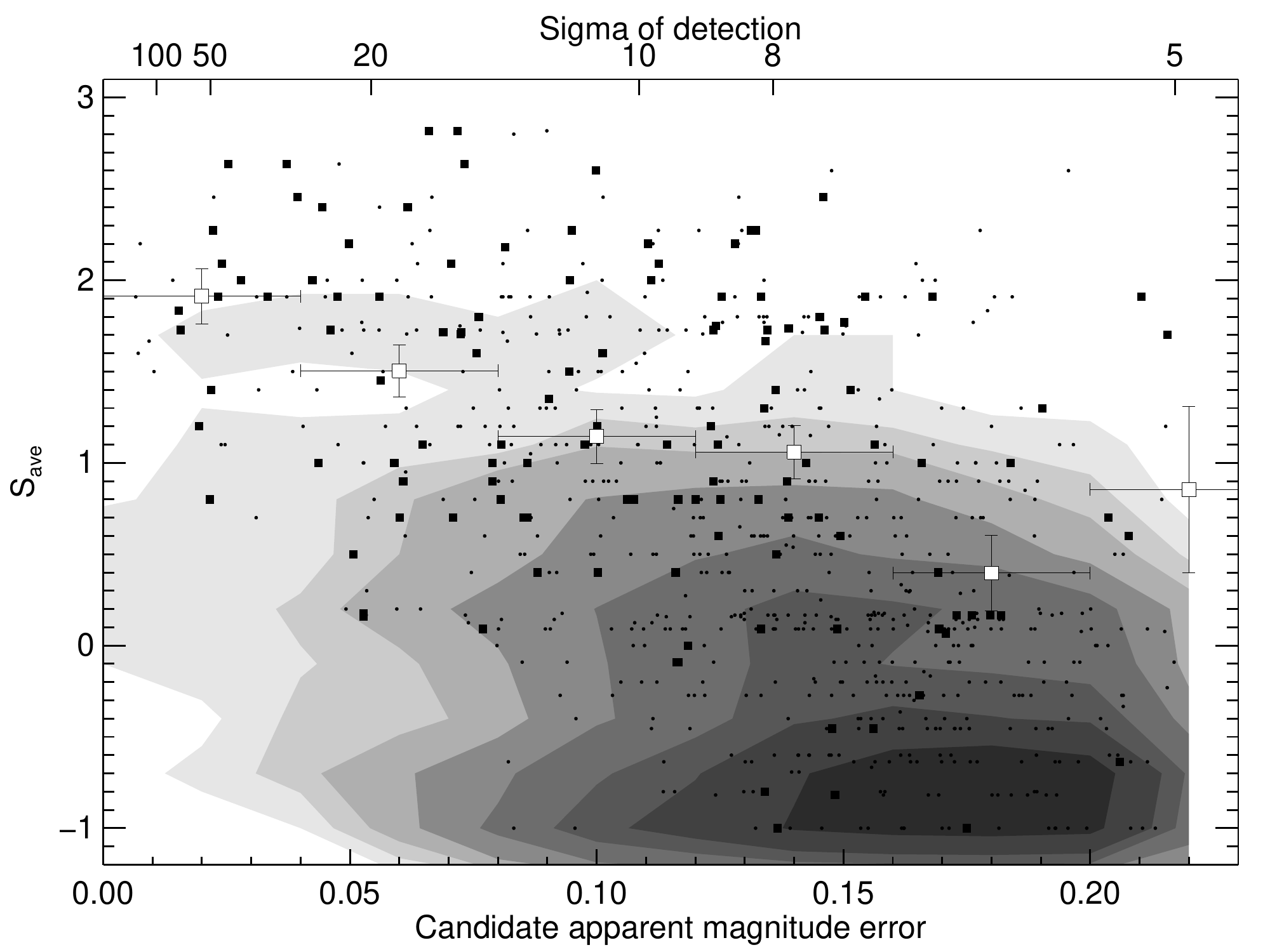}
\caption{The Galaxy Zoo Supernova scores $S_{\mathrm{ave}}$ of PTF
  candidates of various types as a function of their apparent $R$
  detection magnitude (left) and the error in that magnitude (right).
  The filled circles show PTF objects believed to be SN-like
  transients, filled squares show the confirmed SNe, while the
  contours show the distribution of all $\sim14,000$ PTF candidates.
  The open squares show the average SN scores in bins of magnitude (or
  magnitude error). For these candidates, the trend of decreasing
  score with increasing magnitude is significant at about $3\sigma$,
  and with increasing magnitude error at $\sim6\sigma$.  Note that
  only detections of $5\sigma$ significance or greater are uploaded to
  the zoo, hence the cut-off in the right-hand
  panel.\label{fig:magscore}}
\end{figure*}

Fig.~\ref{fig:magscore} shows candidate scores from Galaxy Zoo
Supernovae as a function of the photometric apparent $R$ magnitude of
the candidate (with the host light subtracted) and the magnitude
error, both taken from the P48 PTF search pipeline.  We plot these
relations separately for spectroscopically confirmed SNe and PTF
transients, and show the comparison with all PTF candidates as a set
of contours. This latter comparison highlights just what a small
fraction of all the PTF
candidates the real SNe and transients represent.\\

Fig.~\ref{fig:magscore} shows a few interesting trends. For the
confirmed SNe, there is a mild decrease in $S_{\mathrm{ave}}$ as the
candidates become fainter (or have a larger error), at about
$\sim3\sigma$ significance, or $\sim6\sigma$ when considering the
magnitude error. (There is an equivalent trend for all the PTF
transients.) This is expected -- at fainter magnitudes, SNe become
harder to identify visually with a noisier detection, and the
classification becomes more subjective. The SNe are also likely to be
at higher redshift, and thus perhaps appear more centrally located in
fainter host galaxies and are more likely to fail the final step in
the decision tree (Fig.~\ref{fig:decision_tree}).
\\\\
Nonetheless, Galaxy Zoo Supernovae clearly identifies and scores
highly the bulk of the SNe from PTF, and at bright to intermediate
magnitudes the separation of SNe is robust. Even at fainter
magnitudes, the majority of the SNe score $S_{\mathrm{ave}}>0$, and
above a detection significance of $\sim8\sigma$, the zoo scores all
SNe and the vast majority of PTF transients at $S_{\mathrm{ave}}>0$.

\subsection{The scoring model}
\label{sec:scoring-model}
\begin{figure}
\includegraphics[width=0.49\textwidth]{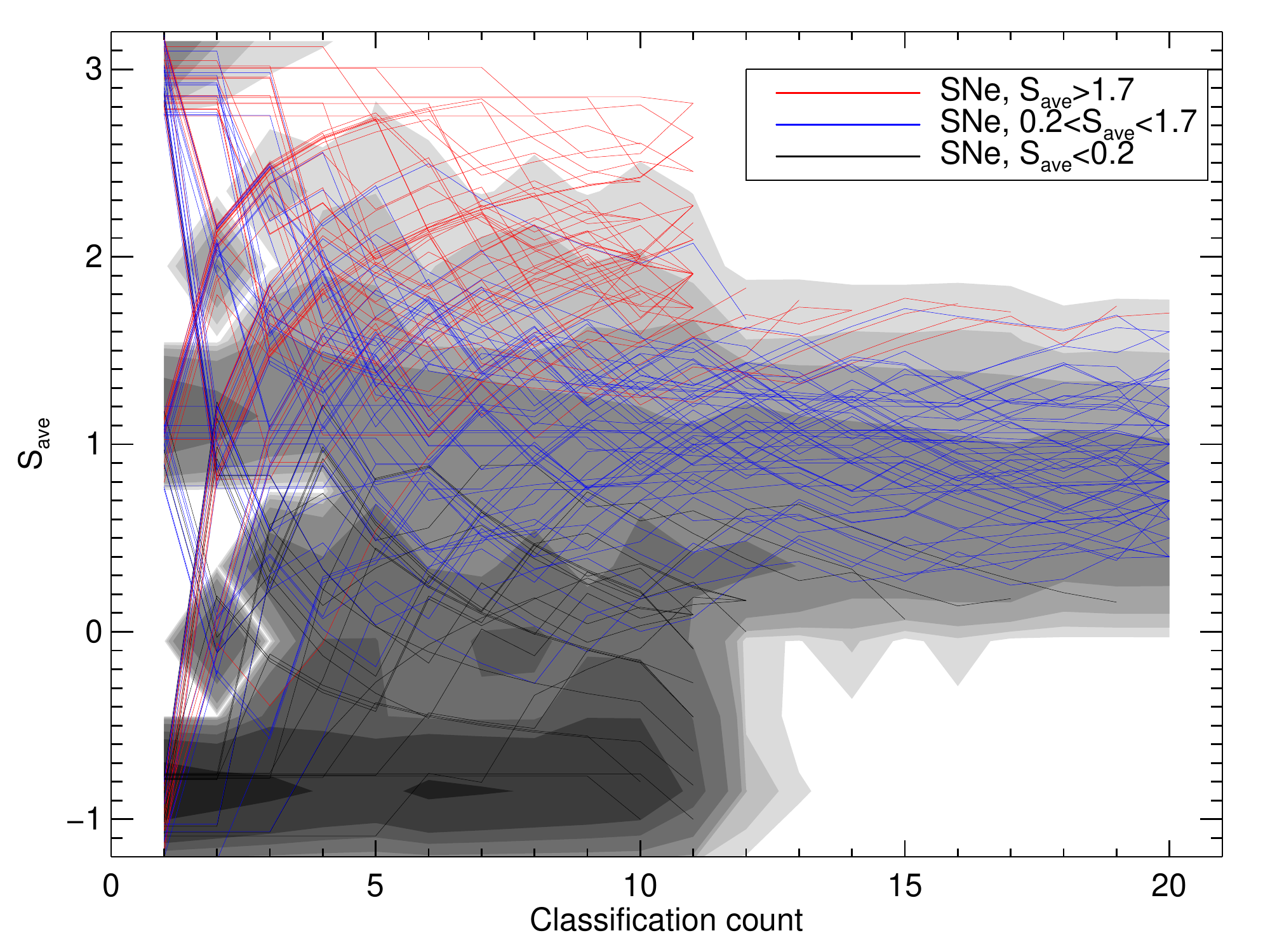}
\caption{The Galaxy Zoo Supernova scores $S_{\mathrm{ave}}$ of PTF
  candidates of various types as a function of the number of
  classifications they have received. Each line represents a
  spectroscopically confirmed PTF SN. Those in red have a final
  $S_{\mathrm{ave}}>1.7$, those in black a final
  $S_{\mathrm{ave}}<0.2$, and those in blue intermediate scores. Only
  candidates scored with $0.2<S_{\mathrm{ave}}<1.7$ continue to be
  classified beyond 10 classifications. The grey contours show the
  trajectories of all PTF candidates that were classified by the zoo,
  regardless of any spectroscopic typing. As the scores as highly
  quantised (each classification can only result in a score of -1, 1
  or 3), each line representing a PTF SN is offset slightly in
  $S_{\mathrm{ave}}$ for clarity.\label{fig:scoretraj}}
\end{figure}

An analysis of the scoring model can reveal optimisations that can be
made to the number of classifications required for each candidate. As
an example, we show the `trajectory' of $S_{\mathrm{ave}}$ for PTF
candidates as a function of the number of classifications in
Fig.~\ref{fig:scoretraj}. As expected, the variation in
$S_{\mathrm{ave}}$ when adding additional classifications is larger
when the total number of classifications is small compared to when
many classifications are available. It is also apparent that once
$\sim15$ classifications have been received, very few candidates
change $S_{\mathrm{ave}}$ significantly.\\

\begin{figure}
\includegraphics[width=0.49\textwidth]{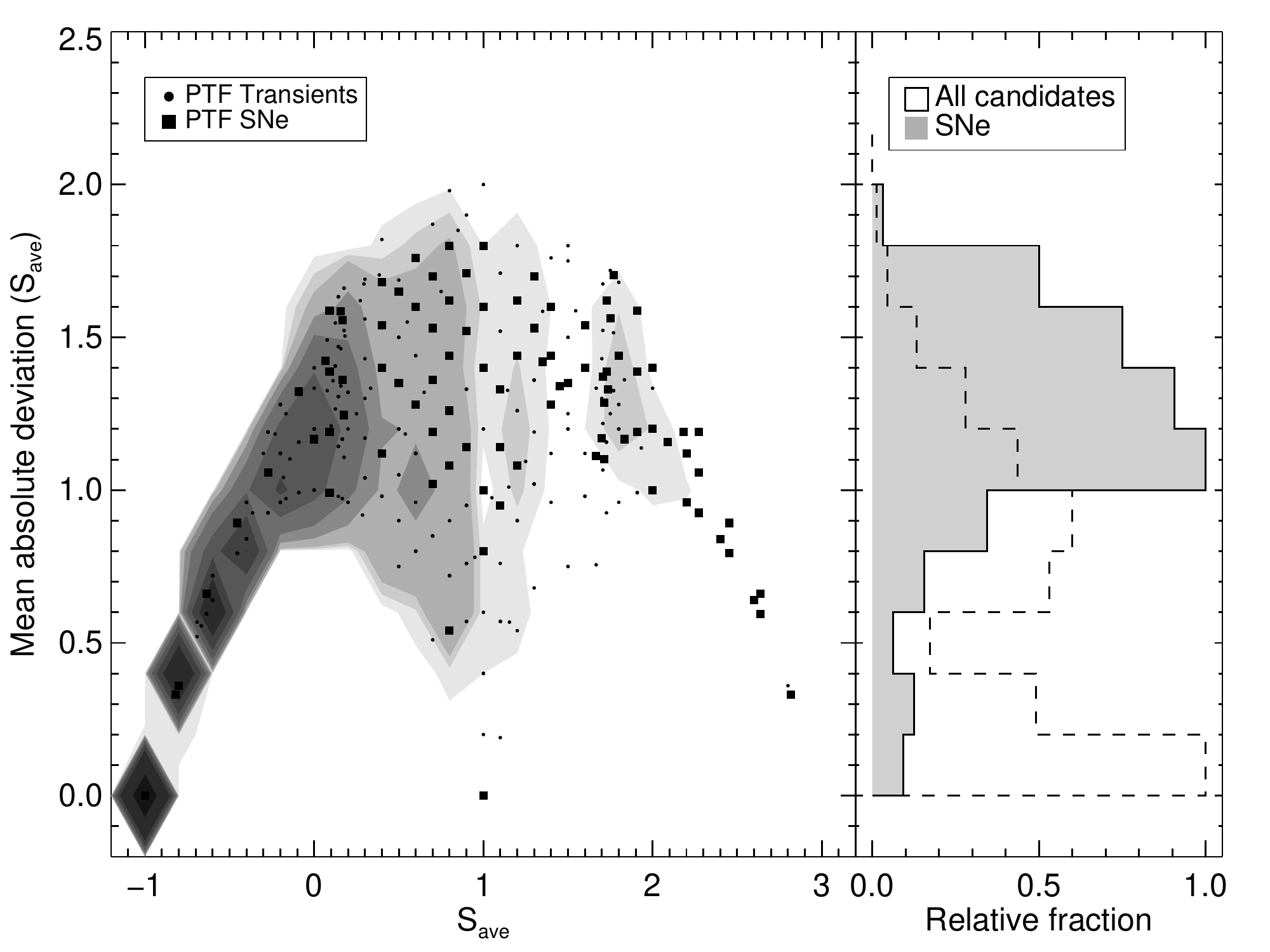}
\caption{The dispersion (mean absolute deviation) of the Galaxy Zoo
  Supernova scores as a function of the scores themselves
  ($S_{\mathrm{ave}}$) for PTF candidates of various types. The filled
  circles show PTF objects believed to be SN-like transients, filled
  squares show the confirmed SNe, while the contours show the
  distribution of all $\sim14,000$ PTF candidates. The histograms show
  the distribution of the mean absolute deviations in
  $S_{\mathrm{ave}}$ for confirmed SNe and for all PTF candidates. In
  principle, with a more refined scoring grid the dispersion in each
  candidate score could be used as a measure of the confidence of the
  final classification, but our current simple decision tree does not
  permit this.\label{fig:scoredev}}
\end{figure}

We also examine the dispersion in each of the Galaxy Zoo Supernova
scores, as calculated from the individual classifications, as a
function of the scores themselves. Fig.~\ref{fig:scoredev} plots the
mean absolute deviation in the score of each classified candidate as a
function of the final candidate score. As the individual scores from
which each $S_{\mathrm{ave}}$ is calculated are highly quantised (each
classification can only result in a score of -1, 1 or 3), the
resulting plot is highly structured. In particular, objects with
$S_{\mathrm{ave}}$ of -1 or 3 must have a dispersion of zero, and a
further dip in the dispersion is also seen around the third scoring
possibility, 1. While in principle the dispersion in the score might
be thought of as a good measure of the classification confidence
(measuring, in essence, the agreement between individual classifiers),
the current simple decision tree is not refined enough to allow this
statistic to be useful.\\

There therefore exists some room to improve the scoring model used by
Galaxy Zoo Supernovae (and hence the efficiency of the project). A
detailed analysis of the data in Fig.~\ref{fig:scoretraj} shows that
reducing the number of classifications needed before a candidate is
considered classified ($\S$~\ref{sec:scoring}) from $>20$ to $>15$
(for intermediate scoring events) and from $>10$ to $>8$ (for low and
high scoring events) would reduce the total number of classifications
recorded by $\sim20 \%$, while only moving a handful of candidates
($<5$\%) across the boundaries of $S_{\mathrm{ave}}=1.7$ and
$S_{\mathrm{ave}}=0.2$.
\\\\
In principle, an analysis of which volunteers consistently get the
classifications correct (when compared to a professional astronomer or
a spectroscopic classification) could be used to weight different
volunteer responses. For example, an experienced classifier with a
consistent history of correct responses could have a larger weight
than a novice volunteer -- there is evidence from
Fig.~\ref{fig:scoretraj} that even good SN candidates can receive the
lowest score (many SN trajectories start at an $S_{\mathrm{ave}}=-1$).
Such a feature is not yet implemented in Galaxy Zoo Supernovae, but
could be used to arrive at a final $S_{\mathrm{ave}}$ more quickly.

\subsection{Volunteer behaviour}
\label{sec:user-behaviour}
To date, over 13,000 individuals from the Zooniverse community have
visited the Galaxy Zoo Supernovae site and 2,800 have classified one or more SN
candidates.  This project relies upon the rapid classification of SN
candidates and although the community is relatively small compared to
e.g. Galaxy Zoo, a combination of email alerts and a committed core of
a few hundred individuals has made Galaxy Zoo Supernovae a success.
\\\\
An analysis of the fraction of classifications contributed compared to
the average number of classifications per user shows that close to
90\% of the classifications in Galaxy Zoo Supernovae are contributed
by less than 20\% of the community. In the Galaxy Zoo 2 project
\citep[][Lintott et al., in prep.]{2010arXiv1003.0449M}, close to 50\% of the classifications were by individuals
whose total classification count was less than 10 galaxies; for Galaxy
Zoo Supernovae that fraction is 3\%.

\section{Future directions}
\label{sec:future}

This paper has introduced Galaxy Zoo Supernovae, a new web-based
citizen science project modelled after `Galaxy Zoo', that uses members
of the public to identify good supernova candidates from wide-field
imaging data. Using data from the Palomar Transient Factory (PTF), we
have shown that the citizen scientists are extremely good at
identifying real SNe from amongst the thousands of candidates that PTF
generates, with only a small `false negative' rate at the faintest
candidate magnitudes.
\\\\
Clearly, Galaxy Zoo Supernovae is not restricted to PTF data and can
in principle be applied to any future imaging survey, such as
SkyMapper \citep[e.g.,][]{2007PASA...24....1K} or Pan-STARRS-1
\citep[e.g.,][]{2004SPIE.5489...11K}. The candidate upload mechanism
is flexible, and the triplet format (Fig.~\ref{fig:egtriplets})
simple, with custom results pages easily produced for individual
surveys. Perhaps the most exciting aspect for massive future transient
surveys such as the Large Synoptic Survey Telescope (LSST) will be the
use of Galaxy Zoo Supernovae classification data to improve the
training and accuracy of automated machine-learning transient
classifiers \citep{2009ASPC..411..493S}.
\\\\
The underlying concept of Galaxy Zoo Supernovae is easily extended.
For example, there is also no need to restrict the project to single
images of new transient events. Multiple images of a potential SN from
different epochs, i.e. a candidate history, could also be uploaded to
improve the accuracy of the classifications and thus reduce the possibility of a mis-classification due to a single poor subtraction.  If this included data
from before the candidate was first detected, those candidates with a
history of poor subtractions could quite trivially be eliminated.
Those asteroids and moving objects which do get uploaded could also be
removed by visually comparing the candidate position on several
epochs. Galaxy Zoo Supernovae could also be used to identify new
transients triggered by detections at other wavelengths, for example
to quickly identify optical counterparts to gamma-ray bursts, where
previous optical reference images might not exist and a timely search
is critical for follow-up.
\\\\
Galaxy Zoo Supernovae could also be used for precise volumetric SN (or
any transient) rate determinations. In these calculations, the
efficiency of the search (the ratio of recovered to actual SN events)
needs to be accurately known, as a function of apparent magnitude and
other SN properties. By uploading `fake' candidates (artificial SN
events inserted into the images) as well as real SNe, the reliability
of the zoo can be determined accurately and allow the discovery rate
to be converted into a real physical SN rate.
\\\\
With the discovery stream of new transient types becoming ever larger,
and the dramatic increase set to continue with future surveys such as
the LSST, the burden of identifying the best new candidates increases
correspondingly. By engaging the considerable interest and enthusiasm
of the public, we have demonstrated that citizen science projects like
Galaxy Zoo Supernovae can play a major role in ongoing and future
transient surveys.

\section*{Acknowledgements}

We acknowledge the valuable contributions of the Zooniverse community
without which this project would not have been possible. AS
acknowledges support from the Leverhulme Trust.  MS acknowledges
support from the Royal Society.  MS and AG acknowledge support from a
Weizmann--UK ``Making conenctions'' grant. CJL acknowledges support
from the STFC Science in Society Program and The Leverhulme Trust.
PEN acknowledges support from the US Department of Energy Scientific
Discovery through Advanced Computing program under contract
DE-FG02-06ER06-04. KS acknowledges support from a NASA Einstein
Postdoctoral Fellowship grant number PF9-00069, issued by the Chandra
X-ray Observatory Center, which is operated by the Smithsonian
Astrophysical Observatory for and on behalf of NASA under contract
NAS8-03060. JSB acknowledges support of an NSF-CDI grant ``Real-time
Classification of Massive Time-series Data Streams'' (Award \#0941742).
The National Energy Research Scientific Computing Center, which is
supported by the Office of Science of the U.S. Department of Energy
under Contract No.  DE-AC02-05CH11231, provided staff, computational
resources and data storage for this project.
\\\\
The William Herschel Telescope is operated on the island of La Palma
by the Isaac Newton Group in the Spanish Observatorio del Roque de los
Muchachos of the Instituto de Astrofísica de Canarias.  Observations
obtained with the Samuel Oschin Telescope at the Palomar Observatory
as part of the Palomar Transient Factory project: a scientific
collaboration between the California Institute of Technology, Columbia
University, Las Cumbres Observatory, the Lawrence Berkeley National
Laboratory, the National Energy Research Scientific Computing Center,
the University of Oxford, and the Weizmann Institute of Science.

\bibliographystyle{mn2e}

\begin{thebibliography}{17}
\expandafter\ifx\csname natexlab\endcsname\relax\def\natexlab#1{#1}\fi

\bibitem[{{Alard}(2000)}]{2000A&AS..144..363A}
{Alard} C., 2000, \aaps, 144, 363

\bibitem[{{Astier} {et~al.}(2006){Astier}, {Guy}, {Regnault}, {Pain},
  {Aubourg}, {Balam}, {Basa}, {Carlberg}, {Fabbro}, {Fouchez}, {Hook},
  {Howell}, {Lafoux}, {Neill}, {Palanque-Delabrouille}, {Perrett}, {Pritchet},
  {Rich}, {Sullivan}, {Taillet}, {Aldering}, {Antilogus}, {Arsenijevic},
  {Balland}, {Baumont}, {Bronder}, {Courtois}, {Ellis}, {Filiol}, {Gon{\c
  c}alves}, {Goobar}, {Guide}, {Hardin}, {Lusset}, {Lidman}, {McMahon},
  {Mouchet}, {Mourao}, {Perlmutter}, {Ripoche}, {Tao}, \&
  {Walton}}]{2006A&A...447...31A}
{Astier} P., {Guy} J., {Regnault} N., {Pain} R., {Aubourg} E., {Balam} D.,
  {Basa} S., {Carlberg} R.~G., {Fabbro} S., {Fouchez} D., {Hook} I.~M.,
  {Howell} D.~A., {Lafoux} H., {Neill} J.~D., {Palanque-Delabrouille} N.,
  {Perrett} K., {Pritchet} C.~J., {Rich} J., {Sullivan} M., {Taillet} R.,
  {Aldering} G., {Antilogus} P., {Arsenijevic} V., {Balland} C., {Baumont} S.,
  {Bronder} J., {Courtois} H., {Ellis} R.~S., {Filiol} M., {Gon{\c c}alves}
  A.~C., {Goobar} A., {Guide} D., {Hardin} D., {Lusset} V., {Lidman} C.,
  {McMahon} R., {Mouchet} M., {Mourao} A., {Perlmutter} S., {Ripoche} P., {Tao}
  C., {Walton} N., 2006, \aap, 447, 31

\bibitem[{{Bertin} \& {Arnouts}(1996)}]{1996A&AS..117..393B}
{Bertin} E., {Arnouts} S., 1996, \aaps, 117, 393

\bibitem[{{Frieman} {et~al.}(2008){Frieman}, {Bassett}, {Becker}, {Choi},
  {Cinabro}, {DeJongh}, {Depoy}, {Dilday}, {Doi}, {Garnavich}, {Hogan},
  {Holtzman}, {Im}, {Jha}, {Kessler}, {Konishi}, {Lampeitl}, {Marriner},
  {Marshall}, {McGinnis}, {Miknaitis}, {Nichol}, {Prieto}, {Riess}, {Richmond},
  {Romani}, {Sako}, {Schneider}, {Smith}, {Takanashi}, {Tokita}, {van der
  Heyden}, {Yasuda}, {Zheng}, {Adelman-McCarthy}, {Annis}, {Assef},
  {Barentine}, {Bender}, {Blandford}, {Boroski}, {Bremer}, {Brewington},
  {Collins}, {Crotts}, {Dembicky}, {Eastman}, {Edge}, {Edmondson}, {Elson},
  {Eyler}, {Filippenko}, {Foley}, {Frank}, {Goobar}, {Gueth}, {Gunn},
  {Harvanek}, {Hopp}, {Ihara}, {Ivezi{\'c}}, {Kahn}, {Kaplan}, {Kent},
  {Ketzeback}, {Kleinman}, {Kollatschny}, {Kron}, {Krzesi{\'n}ski}, {Lamenti},
  {Leloudas}, {Lin}, {Long}, {Lucey}, {Lupton}, {Malanushenko}, {Malanushenko},
  {McMillan}, {Mendez}, {Morgan}, {Morokuma}, {Nitta}, {Ostman}, {Pan},
  {Rockosi}, {Romer}, {Ruiz-Lapuente}, {Saurage}, {Schlesinger}, {Snedden},
  {Sollerman}, {Stoughton}, {Stritzinger}, {Subba Rao}, {Tucker}, {Vaisanen},
  {Watson}, {Watters}, {Wheeler}, {Yanny}, \& {York}}]{2008AJ....135..338F}
{Frieman} J.~A., {Bassett} B., {Becker} A., {Choi} C., {Cinabro} D., {DeJongh}
  F., {Depoy} D.~L., {Dilday} B., {Doi} M., {Garnavich} P.~M., {Hogan} C.~J.,
  {Holtzman} J., {Im} M., {Jha} S., {Kessler} R., {Konishi} K., {Lampeitl} H.,
  {Marriner} J., {Marshall} J.~L., {McGinnis} D., {Miknaitis} G., {Nichol}
  R.~C., {Prieto} J.~L., {Riess} A.~G., {Richmond} M.~W., {Romani} R., {Sako}
  M., {Schneider} D.~P., {Smith} M., {Takanashi} N., {Tokita} K., {van der
  Heyden} K., {Yasuda} N., {Zheng} C., {Adelman-McCarthy} J., {Annis} J.,
  {Assef} R.~J., {Barentine} J., {Bender} R., {Blandford} R.~D., {Boroski}
  W.~N., {Bremer} M., {Brewington} H., {Collins} C.~A., {Crotts} A., {Dembicky}
  J., {Eastman} J., {Edge} A., {Edmondson} E., {Elson} E., {Eyler} M.~E.,
  {Filippenko} A.~V., {Foley} R.~J., {Frank} S., {Goobar} A., {Gueth} T.,
  {Gunn} J.~E., {Harvanek} M., {Hopp} U., {Ihara} Y., {Ivezi{\'c}} {\v Z}.,
  {Kahn} S., {Kaplan} J., {Kent} S., {Ketzeback} W., {Kleinman} S.~J.,
  {Kollatschny} W., {Kron} R.~G., {Krzesi{\'n}ski} J., {Lamenti} D., {Leloudas}
  G., {Lin} H., {Long} D.~C., {Lucey} J., {Lupton} R.~H., {Malanushenko} E.,
  {Malanushenko} V., {McMillan} R.~J., {Mendez} J., {Morgan} C.~W., {Morokuma}
  T., {Nitta} A., {Ostman} L., {Pan} K., {Rockosi} C.~M., {Romer} A.~K.,
  {Ruiz-Lapuente} P., {Saurage} G., {Schlesinger} K., {Snedden} S.~A.,
  {Sollerman} J., {Stoughton} C., {Stritzinger} M., {Subba Rao} M., {Tucker}
  D., {Vaisanen} P., {Watson} L.~C., {Watters} S., {Wheeler} J.~C., {Yanny} B.,
  {York} D., 2008, \aj, 135, 338

\bibitem[{{Hillebrandt} \& {Niemeyer}(2000)}]{2000ARA&A..38..191H}
{Hillebrandt} W., {Niemeyer} J.~C., 2000, \araa, 38, 191

\bibitem[{{Kaiser}(2004)}]{2004SPIE.5489...11K}
{Kaiser} N., 2004, in Presented at the Society of Photo-Optical Instrumentation
  Engineers (SPIE) Conference, Vol. 5489, Society of Photo-Optical
  Instrumentation Engineers (SPIE) Conference Series, {J.~M.~Oschmann Jr.},
  ed., pp. 11--22

\bibitem[{{Keller} {et~al.}(2007){Keller}, {Schmidt}, {Bessell}, {Conroy},
  {Francis}, {Granlund}, {Kowald}, {Oates}, {Martin-Jones}, {Preston},
  {Tisserand}, {Vaccarella}, \& {Waterson}}]{2007PASA...24....1K}
{Keller} S.~C., {Schmidt} B.~P., {Bessell} M.~S., {Conroy} P.~G., {Francis} P.,
  {Granlund} A., {Kowald} E., {Oates} A.~P., {Martin-Jones} T., {Preston} T.,
  {Tisserand} P., {Vaccarella} A., {Waterson} M.~F., 2007, Publications of the
  Astronomical Society of Australia, 24, 1

\bibitem[{{Law} {et~al.}(2009){Law}, {Kulkarni}, {Dekany}, {Ofek}, {Quimby},
  {Nugent}, {Surace}, {Grillmair}, {Bloom}, {Kasliwal}, {Bildsten}, {Brown},
  {Cenko}, {Ciardi}, {Croner}, {Djorgovski}, {van Eyken}, {Filippenko}, {Fox},
  {Gal-Yam}, {Hale}, {Hamam}, {Helou}, {Henning}, {Howell}, {Jacobsen},
  {Laher}, {Mattingly}, {McKenna}, {Pickles}, {Poznanski}, {Rahmer}, {Rau},
  {Rosing}, {Shara}, {Smith}, {Starr}, {Sullivan}, {Velur}, {Walters}, \&
  {Zolkower}}]{2009PASP..121.1395L}
{Law} N.~M., {Kulkarni} S.~R., {Dekany} R.~G., {Ofek} E.~O., {Quimby} R.~M.,
  {Nugent} P.~E., {Surace} J., {Grillmair} C.~C., {Bloom} J.~S., {Kasliwal}
  M.~M., {Bildsten} L., {Brown} T., {Cenko} S.~B., {Ciardi} D., {Croner} E.,
  {Djorgovski} S.~G., {van Eyken} J., {Filippenko} A.~V., {Fox} D.~B.,
  {Gal-Yam} A., {Hale} D., {Hamam} N., {Helou} G., {Henning} J., {Howell}
  D.~A., {Jacobsen} J., {Laher} R., {Mattingly} S., {McKenna} D., {Pickles} A.,
  {Poznanski} D., {Rahmer} G., {Rau} A., {Rosing} W., {Shara} M., {Smith} R.,
  {Starr} D., {Sullivan} M., {Velur} V., {Walters} R., {Zolkower} J., 2009,
  \pasp, 121, 1395

\bibitem[{{Lintott} {et~al.}(2010){Lintott}, {Schawinski}, {Bamford}, {Slosar},
  {Land}, {Thomas}, {Edmondson}, {Masters}, {Nichol}, {Raddick}, {Szalay},
  {Andreescu}, {Murray}, \& {Vandenberg}}]{2010arXiv1007.3265L}
{Lintott} C., {Schawinski} K., {Bamford} S., {Slosar} A., {Land} K., {Thomas}
  D., {Edmondson} E., {Masters} K., {Nichol} R., {Raddick} J., {Szalay} A.,
  {Andreescu} D., {Murray} P., {Vandenberg} J., 2010, ArXiv e-prints

\bibitem[{{Lintott} {et~al.}(2008){Lintott}, {Schawinski}, {Slosar}, {Land},
  {Bamford}, {Thomas}, {Raddick}, {Nichol}, {Szalay}, {Andreescu}, {Murray}, \&
  {Vandenberg}}]{2008MNRAS.389.1179L}
{Lintott} C.~J., {Schawinski} K., {Slosar} A., {Land} K., {Bamford} S.,
  {Thomas} D., {Raddick} M.~J., {Nichol} R.~C., {Szalay} A., {Andreescu} D.,
  {Murray} P., {Vandenberg} J., 2008, \mnras, 389, 1179

\bibitem[{{Masters} {et~al.}(2010){Masters}, {Nichol}, {Hoyle}, {Lintott},
  {Bamford}, {Edmondson}, {Fortson}, {Keel}, {Schawinski}, {Smith}, \&
  {Thomas}}]{2010arXiv1003.0449M}
{Masters} K.~L., {Nichol} R.~C., {Hoyle} B., {Lintott} C., {Bamford} S.,
  {Edmondson} E.~M., {Fortson} L., {Keel} W.~C., {Schawinski} K., {Smith} A.,
  {Thomas} D., 2010, ArXiv e-prints

\bibitem[{{Nugent} {et~al.}(2010){Nugent}, {Cenko}, {Miller}, {Poznanski},
  {Bloom}, {Filippenko}, {Sullivan}, {Howell}, {Quimby}, {Ofek}, {Kasliwal},
  {Kulkarni}, {Law}, {Dekany}, {Rahmer}, {Hale}, {Smith}, {Zolkower}, {Velur},
  {Walters}, {Henning}, {Bui}, {McKenna}, \& {Jacobsen}}]{2010ATel.2600....1N}
{Nugent} P., {Cenko} S.~B., {Miller} A.~M., {Poznanski} D., {Bloom} J.~S.,
  {Filippenko} A.~V., {Sullivan} M., {Howell} D.~A., {Quimby} R.~M., {Ofek}
  E.~O., {Kasliwal} M.~M., {Kulkarni} S.~R., {Law} N.~M., {Dekany} R.~G.,
  {Rahmer} G., {Hale} D., {Smith} R., {Zolkower} J., {Velur} V., {Walters} R.,
  {Henning} J., {Bui} K., {McKenna} D., {Jacobsen} J., 2010, The Astronomer's
  Telegram, 2600, 1

\bibitem[{{Perrett} {et~al.}(2010){Perrett}, {Balam}, {Sullivan}, {Pritchet},
  {Conley}, {Carlberg}, {Astier}, {Balland}, {Basa}, {Fouchez}, {Guy},
  {Hardin}, {Hook}, {Howell}, {Pain}, \& {Regnault}}]{2010AJ....140..518P}
{Perrett} K., {Balam} D., {Sullivan} M., {Pritchet} C., {Conley} A., {Carlberg}
  R., {Astier} P., {Balland} C., {Basa} S., {Fouchez} D., {Guy} J., {Hardin}
  D., {Hook} I.~M., {Howell} D.~A., {Pain} R., {Regnault} N., 2010, \aj, 140,
  518

\bibitem[{{Rau} {et~al.}(2009){Rau}, {Kulkarni}, {Law}, {Bloom}, {Ciardi},
  {Djorgovski}, {Fox}, {Gal-Yam}, {Grillmair}, {Kasliwal}, {Nugent}, {Ofek},
  {Quimby}, {Reach}, {Shara}, {Bildsten}, {Cenko}, {Drake}, {Filippenko},
  {Helfand}, {Helou}, {Howell}, {Poznanski}, \&
  {Sullivan}}]{2009PASP..121.1334R}
{Rau} A., {Kulkarni} S.~R., {Law} N.~M., {Bloom} J.~S., {Ciardi} D.,
  {Djorgovski} G.~S., {Fox} D.~B., {Gal-Yam} A., {Grillmair} C.~C., {Kasliwal}
  M.~M., {Nugent} P.~E., {Ofek} E.~O., {Quimby} R.~M., {Reach} W.~T., {Shara}
  M., {Bildsten} L., {Cenko} S.~B., {Drake} A.~J., {Filippenko} A.~V.,
  {Helfand} D.~J., {Helou} G., {Howell} D.~A., {Poznanski} D., {Sullivan} M.,
  2009, \pasp, 121, 1334

\bibitem[{{Sako} {et~al.}(2008){Sako}, {Bassett}, {Becker}, {Cinabro},
  {DeJongh}, {Depoy}, {Dilday}, {Doi}, {Frieman}, {Garnavich}, {Hogan},
  {Holtzman}, {Jha}, {Kessler}, {Konishi}, {Lampeitl}, {Marriner}, {Miknaitis},
  {Nichol}, {Prieto}, {Riess}, {Richmond}, {Romani}, {Schneider}, {Smith},
  {Subba Rao}, {Takanashi}, {Tokita}, {van der Heyden}, {Yasuda}, {Zheng},
  {Barentine}, {Brewington}, {Choi}, {Dembicky}, {Harnavek}, {Ihara}, {Im},
  {Ketzeback}, {Kleinman}, {Krzesi{\'n}ski}, {Long}, {Malanushenko},
  {Malanushenko}, {McMillan}, {Morokuma}, {Nitta}, {Pan}, {Saurage}, \&
  {Snedden}}]{2008AJ....135..348S}
{Sako} M., {Bassett} B., {Becker} A., {Cinabro} D., {DeJongh} F., {Depoy}
  D.~L., {Dilday} B., {Doi} M., {Frieman} J.~A., {Garnavich} P.~M., {Hogan}
  C.~J., {Holtzman} J., {Jha} S., {Kessler} R., {Konishi} K., {Lampeitl} H.,
  {Marriner} J., {Miknaitis} G., {Nichol} R.~C., {Prieto} J.~L., {Riess} A.~G.,
  {Richmond} M.~W., {Romani} R., {Schneider} D.~P., {Smith} M., {Subba Rao} M.,
  {Takanashi} N., {Tokita} K., {van der Heyden} K., {Yasuda} N., {Zheng} C.,
  {Barentine} J., {Brewington} H., {Choi} C., {Dembicky} J., {Harnavek} M.,
  {Ihara} Y., {Im} M., {Ketzeback} W., {Kleinman} S.~J., {Krzesi{\'n}ski} J.,
  {Long} D.~C., {Malanushenko} E., {Malanushenko} V., {McMillan} R.~J.,
  {Morokuma} T., {Nitta} A., {Pan} K., {Saurage} G., {Snedden} S.~A., 2008,
  \aj, 135, 348

\bibitem[{{Smartt}(2009)}]{2009ARA&A..47...63S}
{Smartt} S.~J., 2009, \araa, 47, 63

\bibitem[{{Starr} {et~al.}(2009){Starr}, {Bloom}, {Brewer}, {Butler},
  {Poznanski}, {Rischard}, \& {Klein}}]{2009ASPC..411..493S}
{Starr} D.~L., {Bloom} J.~S., {Brewer} J.~M., {Butler} N.~R., {Poznanski} D.,
  {Rischard} M., {Klein} C., 2009, in Astronomical Society of the Pacific
  Conference Series, Vol. 411, Astronomical Society of the Pacific Conference
  Series, {D.~A.~Bohlender, D.~Durand, \& P.~Dowler}, ed., pp. 493--+

\end{thebibliography}

\end{document}